\documentclass[%
 aip,
 amsmath,amssymb,
 reprint,%
]{revtex4-1}

\usepackage{graphicx}
\usepackage{dcolumn}
\usepackage{bm}

\usepackage[utf8]{inputenc}
\usepackage[T1]{fontenc}
\usepackage{mathptmx}
\usepackage{etoolbox}

\usepackage{orcidlink}
\usepackage{amsmath, amssymb, mathrsfs, mathtools, bm}
\usepackage{algpseudocode}

\usepackage{xcolor}
\usepackage{ulem}
\usepackage{multirow, booktabs}

\makeatletter
\def\@email#1#2{%
 \endgroup
 \patchcmd{\titleblock@produce}
  {\frontmatter@RRAPformat}
  {\frontmatter@RRAPformat{\produce@RRAP{*#1\href{mailto:#2}{#2}}}\frontmatter@RRAPformat}
  {}{}
}%
\makeatother

\allowdisplaybreaks

\newcommand{\Legolas}{\textsc{Legolas}}
\newcommand{\etal}{\textit{et al.}}

\newcommand{\bfb}{\mathbf{B}}
\newcommand{\bfj}{\mathbf{J}}
\newcommand{\bfv}{\mathbf{v}}
\newcommand{\bfk}{\mathbf{k}}

\newcommand{\ey}{\hat{\boldsymbol{e}}_y}
\newcommand{\ez}{\hat{\boldsymbol{e}}_z}
\newcommand{\im}{\mathrm{i}}

\begin{document}

\preprint{AIP/123-QED}

\title[Modification of the resistive tearing instability with Joule heating by shear flow]{Modification of the resistive tearing instability with Joule heating by shear flow}
\author{J. De Jonghe \orcidlink{0000-0003-2443-3903}}
 \email{jkmdj1@st-andrews.ac.uk}
 \affiliation{Centre for mathematical Plasma Astrophysics, KU Leuven, B-3001 Leuven, Belgium}
 \affiliation{School of Mathematics and Statistics, University of St Andrews, St Andrews KY16 9SS, UK}
\author{R. Keppens \orcidlink{0000-0003-3544-2733}}
 \affiliation{Centre for mathematical Plasma Astrophysics, KU Leuven, B-3001 Leuven, Belgium}%

\date{16 February 2024}

\begin{abstract}
We investigate the influence of background shear flow on linear resistive tearing instabilities with Joule heating for two compressible plasma slab configurations: a Harris current sheet and a force-free, shearing magnetic field that varies its direction periodically throughout the slab, possibly resulting in multiple magnetic nullplanes. To do so, we exploit the latest version of the open-source, magnetohydrodynamic spectroscopy tool \Legolas{}. Shear flow is shown to dramatically alter tearing behaviour in the presence of multiple magnetic nullplanes, where the modes become propagating due to the flow. Finally, the tearing growth rate is studied as a function of resistivity, showing where it deviates from analytic scaling laws, as well as the Alfv\'en speed, the plasma-$\beta$, and the velocity parameters, revealing surprising nuance in whether the velocity acts stabilising or destabilising. We show how both slab setups can produce growth rate regimes which deviate from analytic scaling laws, such that systematic numerical spectroscopic studies are truly necessary, for a complete understanding of linear tearing behaviour in flowing plasmas.
\end{abstract}

\maketitle

\section{Introduction}\label{sec:intro}
In many plasma configurations, from astrophysical to laboratory settings, the most dramatic and violent events are often driven by magnetic reconnection. During the reconnection process, magnetic field lines break and reconnect, drastically altering the magnetic topology, e.g. by unravelling a knotted field structure. Consequently, stored magnetic energy is converted into thermal and kinetic energy, accelerating particles and producing matter outflows. It is frequently observed in various phenomena, like coronal mass ejections,\cite{Lorincik2021} the heliospheric current sheet,\cite{Phan2022} and Earth's magnetosphere.\cite{Qi2022}

Due to its importance and ubiquity, reconnection has been a topic of great debate ever since Sweet \cite{Sweet1958} and Parker \cite{Parker1957} first proposed their model, which was later followed by Petschek's model \cite{Petschek1964} to achieve faster reconnection rates in accordance with observations. Irrespective of the specific reconnection process, its initialisation in current sheets is intrinsically linked to tearing instabilities, which occur due to magnetic shear, and create the necessary reconnection points (X-points). These tearing instabilities come in collisional \cite{Furth1963} (resistive) and collisionless \cite{Birn2001} (mediated by the Hall effect and/or electron inertia) varieties, both of which have been studied extensively to understand their role in triggering fast magnetic reconnection.

To reach the fast reconnection regime, the influence of several effects on the resistive tearing mode has been studied for decades. For thin current sheets with a width on the order of the ion inertial length, the importance of the Hall term in the generalised Ohm's law was already shown forty years ago by Terasawa,\cite{Terasawa1983} who showed that it enhances the growth rate of the resistive tearing mode. Following this work, many more studies have honed in on the growth rate modification of resistive tearing in the Hall regime. \cite{Fruchtman1993, Huba2004, Pucci2017, Papini2019, Shi2020, DeJonghe2022} However, the role of the Hall term in magnetic reconnection is not limited to the enhancement of the resistive tearing growth rate. It has also been linked to finite Larmor radius effects, \cite{Wang1993, Kleva1995, Wang2000} it couples to the anisotropic electron pressure tensor, which enhances reconnection rates, \cite{Cai1997, Yin2001} and it can induce a transition to whistler-mediated reconnection. \cite{Mandt1994, Birn2001, Shay2001} Finally, Liu \etal{} \cite{Liu2022} recently showed that the Hall term plays an important role in attaining the proper geometry for fast reconnection.

However, kinetic simulations suggest that the collisionless terms in the generalised Ohm's law do not dominate in collisional plasmas until the current sheet thins to the ion inertial scale, \cite{Bhattacharjee2004, Daughton2009} though collisionless reconnection does occur at larger scales in plasmas with negligible collisionality due to electron inertia.\cite{Coppi1964} Hence, for thicker, sufficiently collisional current sheets, the initial unstable perturbation is expected to be of the resistive tearing variety, with a negligible contribution due to collisionless terms. How these sheets then reduce to the ion inertial scale to establish fast reconnection is not fully understood, though processes like fractal reconnection \cite{Shibata2001} have been suggested.

This then begs the question which other factors might influence the growth rate outside of the Hall regime. It is known that the growth rate is affected by various environmental factors, notably by the background flow. \cite{Li2010, Li2012} In this endeavour, the influence of equilibrium flow on the resistive tearing mode has already been studied extensively using both analytic \cite{Hofmann1975, Pollard1979, Paris1983, Einaudi1986, Chen1990} and numerical \cite{Li2010, Zhang2011, Li2012, Wu2014,Shi2022} methods. Flow's influence is also not limited to the resistive tearing type, as its impact on collisionless tearing has been shown for flow parallel to the reconnecting field \cite{Faganello2010} and to a guide field.\cite{Tassi2014} Furthermore, the role of flow, and particularly flow shear, extends well beyond the modification of tearing growth rates, as it may introduce the Kelvin-Helmholtz instability (KHI) into the system. This instability regularly becomes the dominant instability, \cite{Hofmann1975, Biskamp2000} though it is also observed to co-exist with the plasmoid instability (secondary tearing) in current sheets,\cite{Loureiro2013} at plasma interfaces,\cite{Keppens1999} and in turbulent reconnection.\cite{Borgogno2022}

A profound understanding of instabilities and how to suppress them is also of vital importance in laboratory plasmas, and particularly for fusion research. In fusion devices like tokamaks, tearing instabilities lead to the formation of magnetic islands, which in turn disrupt plasma confinement. There have been many experiments,\cite{Park2013, Shao2021} linear studies,\cite{Chu1995, White2015, Cai2018} and non-linear simulations\cite{Chen1992, Smolyakov2001, Ren2023} studying the effects of both toroidal and poloidal flow. In these torus-like geometries, it is now generally accepted that the toroidal flow shear has a stabilising influence on the system by suppressing island formation.

In this work, we revisit the linear analysis of the resistive tearing mode in the presence of background flow using numerical means. Unlike earlier literature, we eliminate the need for approximations, like incompressibility, by employing the magnetohydrodynamic (MHD) spectroscopic code \Legolas{} \cite[\url{https://legolas.science}]{Claes2020, DeJonghe2022, Claes2023}. With this code we explore plasma stability parametrically for a selection of configurations and show that equilibrium flow can both enhance and suppress the growth rate of the resistive tearing mode depending on the parameter regime. The effect of equilibrium flow has been studied analytically by Chen and Morrison \cite{Chen1990} and we connect our results to the power law predictions from their study.

\section{Setup and conventions}
To study the linear properties of the resistive tearing instability, we linearise the dimensionless, compressible, resistive MHD equations
\begin{align}
	\frac{\partial \rho}{\partial t} = &-\nabla \cdot (\rho \bfv) \,, \label{eq:continuity} \\
	\rho\frac{\partial \bfv}{\partial t} = &-\nabla p - \rho \bfv \cdot \nabla \bfv + \bfj \times \bfb \,,\label{eq:momentum} \\
	\rho\frac{\partial T}{\partial t} = &-\rho \bfv\cdot\nabla T - (\gamma - 1)p\nabla \cdot \bfv + (\gamma - 1)\eta\bfj^2 \,, \label{eq:energy} \\
	\frac{\partial \bfb}{\partial t} = &\nabla \times (\bfv \times \bfb) - \nabla \times (\eta\bfj) \,, \label{eq:induction}
\end{align}
around a one-dimensionally varying equilibrium ($x$-dependence only) and assume Fourier solutions
\begin{equation}\label{eq:fourier}
f_1(\mathbf{r},t) = \hat{f}_1(x)\,\exp\left[ \im\left( k_y y + k_z z - \omega t \right) \right]
\end{equation}
for perturbed quantities $f_1 \in\{\rho_1, \bfv_1, T_1, \bfb_1\}$. Here, $\rho$, $\bfv$, $T$, and $\bfb$ are the usual quantities density, bulk velocity, temperature, and magnetic field, respectively, with subscripts $0$ and $1$ differentiating between equilibrium quantities and perturbations. Furthermore, $p = \rho T$ denotes the pressure, $\bfj = \nabla\times\bfb$ the current density, and $\eta = 10^{-4}$, unless specified otherwise, and $\gamma = 5/3$ are the resistivity and adiabatic index, respectively. Finally, $\bfk = k_y\,\ey + k_z\,\ez$ and $\omega$ are the wave vector and frequency. The resulting algebraic problem is solved for the frequency and corresponding perturbed quantities by \Legolas{}.

Note that our choice of energy equation, Eq. (\ref{eq:energy}), including Joule heating, along with compressibility and a constant resistivity, differs from the isothermal closure used in the original derivation of the tearing mode.\cite{Furth1963}

\subsection{Configurations}
Now consider a semi-infinite plasma confined in the $x$-direction between two perfectly conducting plates, described in Cartesian coordinates. In such a plasma, we examine the resistive tearing instability for two separate equilibrium configurations: a Harris current sheet and a force-free magnetic field.

\subsubsection{Harris current sheet}\label{sec:setup1}
\begin{figure}[h]
    \centering
    \includegraphics[width=0.45\textwidth]{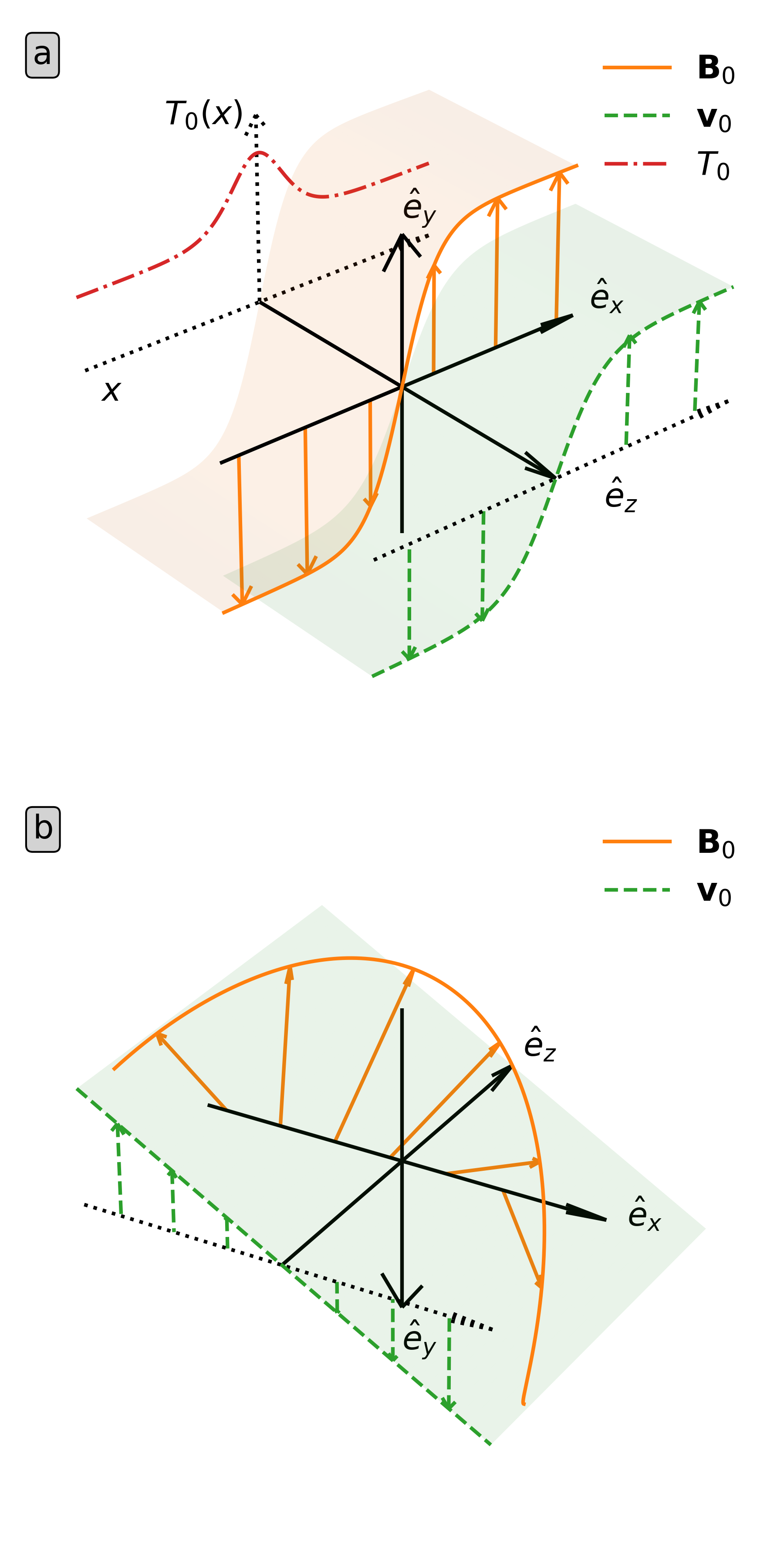}
    \caption{(a) The Harris current sheet equilibrium, Eqs. (\ref{eq:harris-B}-\ref{eq:harris-T}). (b) Equilibrium with the force-free magnetic field, Eqs. (\ref{eq:equil-rotB}). Parameter values were chosen for visual clarity.}
    \label{fig:equilibria}
\end{figure}

For the first configuration, we consider the popular Harris current sheet, as presented in Ref. \onlinecite{Li2010}. In their simulation setup they consider a typical Harris current sheet
\begin{equation}\label{eq:harris-B}
    \bfb_0(x) = B_\mathrm{c} \tanh\left( \frac{x}{a_B} \right)\ \ey,
\end{equation}
which they supplement with a similar velocity profile
\begin{equation}\label{eq:harris-flow}
    \bfv_0(x) = v_\mathrm{c} \tanh\left( \frac{x}{a_v} \right)\ \ey
\end{equation}
and a uniform density $\rho_0 = 1$. The temperature profile is simply obtained by demanding that the total equilibrium pressure (the sum of plasma pressure and magnetic pressure) is constant,\footnote{Sometimes, the temperature is chosen as the constant quantity and subsequently, the density profile is determined by Eq. (\ref{eq:equil-cond}) such as in e.g. Ref. \onlinecite{Goedbloed2019}, Sec. $14.4.1$.} i.e.
\begin{equation}\label{eq:equil-cond}
    \frac{\partial}{\partial x} \left( \rho_0 T_0(x) + \frac{1}{2}\bfb_0^2(x) \right) = 0,
\end{equation}
with a solution
\begin{equation}\label{eq:harris-T}
    T_0(x) = \left( p_\mathrm{c} - \frac{1}{2} \bfb_0^2(x) \right) / \rho_0
\end{equation}
for any constant $p_\mathrm{c}$, set to $p_\mathrm{c} = 1$ throughout this work, unless specified. The profiles are visualised in Fig. \ref{fig:equilibria}(a).

Since $\bfb_0(x)$ vanishes at $x = 0$, so does
\begin{equation}\label{eq:F}
    F(x) = \frac{\bfk}{|\bfk|}\cdot\bfb_0(x)
\end{equation}
for any wave vector $\bfk$, which we choose to be along the $y$-axis, $\bfk = k_y\,\ey$. Hence, the magnetic nullplane (or resonant plane), defined as the location where $F$ vanishes,\cite{Furth1963} is always at $x = 0$ in this configuration, and magnetic reconnection will occur here.

The locations of the perfectly conducting walls at $x_\mathrm{w} = \pm 15\,a_B$ are chosen such that their effect on the tearing instability is negligible (according to Ref. \onlinecite{Ofman1993} the effect of the conducting walls is negligible if they occur at a position $|x_w| \gtrsim 10\,a_B$). Due to the sharp transitions in the equilibrium profiles near the origin and approximately constant behaviour away from the center, the problem is solved using a non-uniform grid concentrated near the origin, as described in App. \ref{app:grid}.

\subsubsection{Force-free magnetic field}\label{sec:setup2}
In the second configuration, the magnetic field strength is fixed whilst its direction varies continuously throughout the plasma slab, similar to the configuration in Refs. \onlinecite{Cross1971, VanHoven1973}. This force-free magnetic field profile is complemented with a constant density and temperature, and a linear velocity profile, as defined in Ref. \onlinecite[Sec. 14.3.3]{Goedbloed2019},
\begin{equation}\label{eq:equil-rotB}
    \begin{aligned}
    &\rho_0(x) = \rho_\mathrm{c}, \qquad &&\bfb_0(x) = \sin(\alpha x)\ \ey + \cos(\alpha x)\ \ez, \\
    &T_0(x) = \frac{\beta_0\bfb_0^2}{2\rho_\mathrm{c}}, &&\bfv_0(x) = v_\mathrm{c} x\ \ey,
    \end{aligned}
\end{equation}
where $\rho_\mathrm{c}$, $v_\mathrm{c}$, and $\alpha$ are constant parameters. As the notation suggests, $\beta_0$ is the equilibrium plasma-$\beta$. Note that $|\bfb_0| = 1$ such that the dimensionless Alfv\'en speed $c_\mathrm{A}$ equals $c_\mathrm{A} = 1/\sqrt{\rho_\mathrm{c}}$. The equilibrium is visualised in Fig. \ref{fig:equilibria}(b).

For this equilibrium we choose $\bfk$ proportional to $\ey$ (and thus parallel to $\bfv_0$), such that there is a magnetic nullplane at $x_0 = 0$, where $F(x_0) = 0$. The perfectly conducting boundaries are placed symmetrically around the nullplane, at $x_\mathrm{w} = \pm 0.5$. Note that additional magnetic nullplanes appear in this interval for sufficiently large $\alpha$.

\subsection{Conventions and technicalities}\label{sec:conventions}
\paragraph{$\psi$-regimes.} In the analytic literature surrounding the resistive tearing instability (notably Refs. \onlinecite{Furth1963, Chen1990}), the mode is usually classified by the behaviour of the normalised, magnetic $x$-perturbation amplitude $\hat{B}_{1x}$, called $\psi$ in Ref. \onlinecite{Furth1963} and subsequent literature, in a resistive layer $[x_0 - \delta, x_0 + \delta]$ around the magnetic nullplane at $x = x_0$, where $F(x_0) = 0$. If $\psi$ is approximately constant across this resistive layer, the mode is called a constant-$\psi$ mode. In general, this corresponds to short wavelengths (large $k$). On the other hand, for longer wavelengths (small $k$), the variation in $\psi$ throughout the resistive layer is not negligible. In this case, the tearing mode is called a nonconstant-$\psi$ mode. This distinction is important because, analytically, the growth rate of the tearing mode scales differently with resistivity $\eta$ for the two regimes. In the static (and small $\bfv_0$) case, the growth rate scales as $\sim\eta^{3/5}$ for constant-$\psi$ modes, whereas for nonconstant-$\psi$ modes, the growth rate scales as $\sim\eta^{1/3}$.\cite{Furth1963,Chen1990,Biskamp2000}

\begin{figure}[h]
    \centering
    \includegraphics[width=0.48\textwidth]{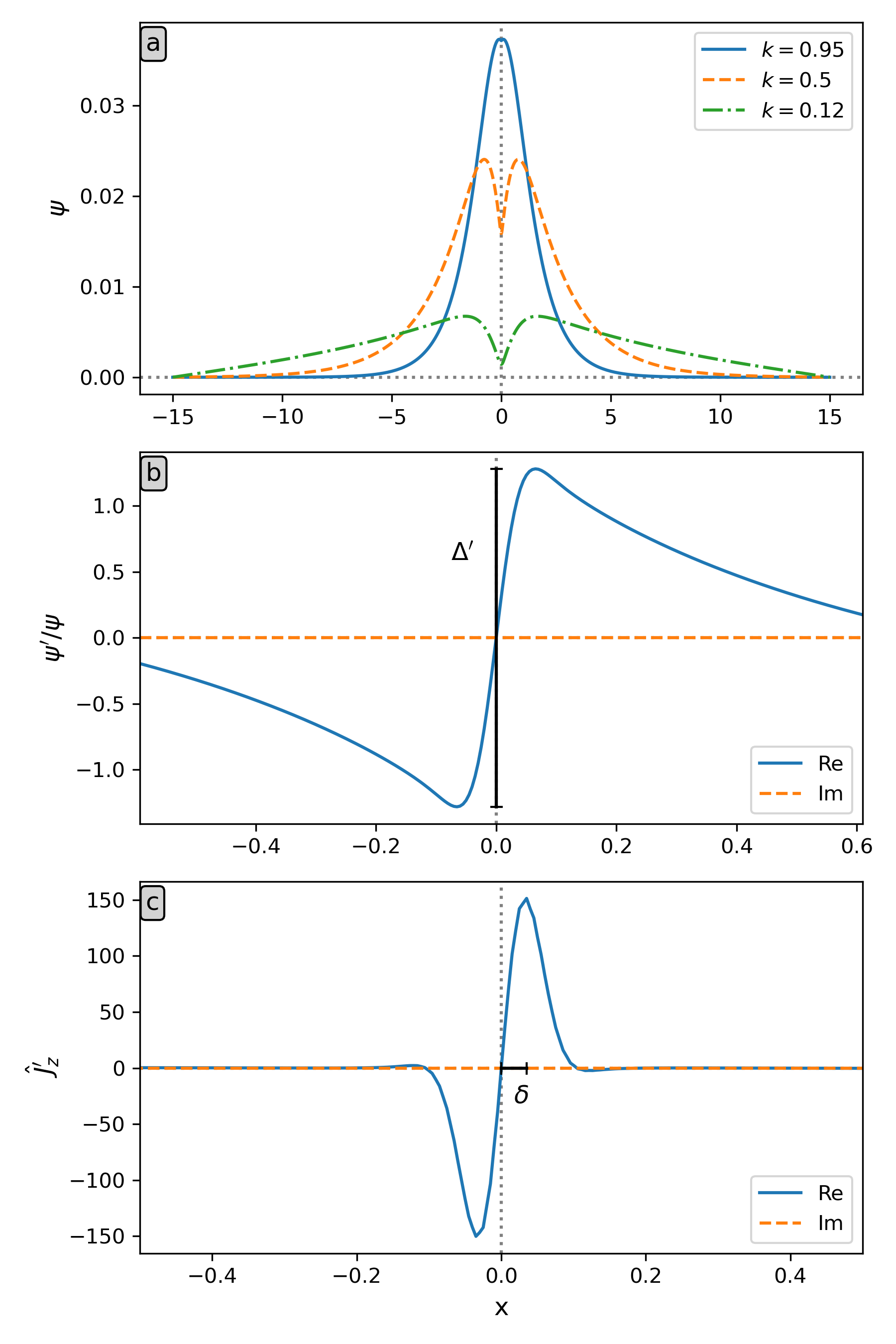}
    \caption{(a) $\psi$ for different wavenumbers in a Harris sheet. (b) $\psi'/\psi$ and (c) $\hat{J}'_z$ for the $k = 0.5$ case from panel (a). $\Delta'$ and $\delta$ indicate the numerical matching quantity and resistive layer halfwidth, respectively.}
    \label{fig:psi-ratio}
\end{figure}

To visualise how $\psi$ changes with $k$, $\psi$ is shown for the tearing mode of the static Harris sheet from Sec. \ref{sec:setup1} ($B_\mathrm{c} = 1$, $a_B = 1$, $v_\mathrm{c} = 0$) for a selection of wavenumbers in Fig. \ref{fig:psi-ratio}(a).

Alongside $\psi$ and the resistive layer halfwidth $\delta$, the literature quantifies the matching quantity\cite{Furth1963}
\begin{equation}
    \Delta' = \left[ \frac{1}{\psi} \frac{\partial\psi}{\partial x} \right]_{x=x_0^-}^{x=x_0^+}.
\end{equation}
In the analytic approach, $\psi$ is obtained by solving the linearised, ideal MHD equations outside of the resistive layer, because resistivity is negligible there, and matching them with the resistive solution inside the layer. However, as an artefact of this approach, the resulting solution has a discontinuity in $\psi'$. To eliminate discontinuities from the calculations, the matching quantity $\Delta'$ is introduced in the analytic approach. Consequently, this quantity appears in the analytic scaling laws, and imposes the condition $\delta |\Delta'| < 1$ for the constant-$\psi$ approximation to hold.\cite{Furth1963,Chen1990,Biskamp2000}

Numerically, $\psi'$ has no true discontinuity. Instead, there is a steep but smooth reversal across the magnetic nullplane. Hence, we here define $\Delta'$ as the difference in $\psi'/\psi$ between the extrema on either side of the nullplane, as seen for the static Harris sheet with $k = 0.5$ in Fig. \ref{fig:psi-ratio}(b). As shown in Ref. \onlinecite{Betar2022}, defining $\delta$ as the distance from the nullplane to the nearest inflexion point of $\hat{J}_z$, i.e. $\hat{J}''_z(\delta) = 0$, as shown in Fig. \ref{fig:psi-ratio}(c), is consistent with boundary layer theory. Though $\hat{J}_z$ is symmetric around the nullplane, we take
\begin{equation}
    \delta = \frac{|x_M - x_m|}{2},
\end{equation}
with $\hat{J}'_z(x_M) = \max(\hat{J}'_z)$ and $\hat{J}'_z(x_m) = \min(\hat{J}'_z)$, to limit the impact of the discretised grid and the numerical differentiation of $\hat{J}_z$. However, if background flow is included, all perturbations become complex. To maintain consistency with the static configurations, the complex factor is chosen in such a way that $\mathrm{Im}(\hat{B}_{1x})$ is positive and symmetric around the nullplane. For $\psi'/\psi$ this factor is eliminated and the real part is used to define $\Delta'$, but this allows us to define $\delta$ based on the now odd function $\mathrm{Re}(\hat{J}'_z)$.

To distinguish between analytic and numerical matching quantities and resistive layer halfwidths, we will refer to them with subscripts $\mathrm{A}$ and $\mathrm{N}$, respectively.

\paragraph{Shear ratio.} Similarly to $F$ in Eq. (\ref{eq:F}), for the equilibrium flow we define the angle-modulated Alfv\'en Mach number (like Ref. \onlinecite{Chen1990})
\begin{equation}\label{eq:G}
    G(x) = \frac{\bfk\cdot\bfv_0(x)}{|\bfk| c_\mathrm{A}}.
\end{equation}
Here, $c_\mathrm{A}$ indicates the dimensionless Alfv\'en speed $c_\mathrm{A} = |\bfb_0|/\sqrt{\rho_0}$. Following Ref. \onlinecite{Chen1990}, the expression
\begin{equation}
R_0 = \left|\frac{G'(x_0)}{F'(x_0)}\right|,
\end{equation}
where the prime denotes the derivative with respect to $x$, acts as a diagnostic parameter to quantify the relative strength of the flow shear compared to the magnetic shear. We will refer to $R_0$ as the shear ratio.

\paragraph{Scaling laws.} Now that we have introduced the $\psi$-regimes and shear ratio $R_0$, we can differentiate between the various conditions under which the growth rate scaling laws were derived as a function of the resistivity $\eta$. The scaling laws are summarised in Table \ref{tab:scaling-laws} as they were obtained by Chen and Morrison.\cite{Chen1990}

\begin{table}[h]
    \centering
    \caption{Analytic scaling laws of the tearing growth rate with the resistivity $\eta$.\cite{Chen1990}}
    \label{tab:scaling-laws}
    \begin{tabular}{lcc}
        \toprule
        \hspace{1.5cm} & Constant-$\psi$ & Nonconstant-$\psi$ \\
        \midrule
        $R_0 \ll 1$ & $\gamma \sim \eta^{3/5}$ & \multirow{2}{*}{$\gamma \sim \eta^{1/3}$} \\
        $R_0 \lesssim 1$ & $\gamma \sim \eta^{1/2}$ & \\
        $R_0 > 1$ & \multicolumn{2}{c}{stabilised} \\
        \bottomrule
    \end{tabular}
\end{table}

\paragraph{Relative quantities.} Throughout this article, we compare the tearing growth rate under the influence of equilibrium flow to the equivalent configuration without flow. In these cases, we opt to use the relative growth rate $\gamma$, which we define as
\begin{equation}
    \gamma = \frac{\text{Im}(\omega_{\text{flow}}) - \text{Im}(\omega_{\text{no flow}})}{\text{Im}(\omega_{\text{no flow}})}.
\end{equation}
Hence, $\gamma$ ranges from $-1$, which means the tearing instability is completely stabilised, to $+\infty$. If $\gamma = 0$, background flow does not alter the growth rate. Similarly, we also define the relative numerical matching quantity
\begin{equation}
    \Delta'_\mathrm{rel} = \frac{\Delta'_\text{N,flow} - \Delta'_\text{N,no flow}}{\Delta'_\text{N,no flow}}.
\end{equation}

\paragraph{Solvers.} Throughout this work, two main solvers are used in \Legolas{}, as described in Ref. \onlinecite{Claes2023}. The first one is the default solver, \texttt{QR-cholesky}, which results in the full spectrum, but is quite slow. The second one is the shift-invert Arnoldi solver, which only computes a selection of modes and is thus faster, but requires a target to converge around. This latter method is preferred for parameter studies once we have an approximation of the growth rate to act as the target.

\section{Results}
In this work, we numerically investigate the flow-sheared resistive tearing mode in two different configurations. In Sec. \ref{sec:gr-harris} we first present visualisations of the perturbed magnetic field and flow in a Harris sheet (see Sec. \ref{sec:setup1}) due to the linear tearing mode, both in the absence and presence of equilibrium flow. Then, we systematically vary the parameters appearing in the shear ratio $R_0$ to evaluate how the growth rate is affected. In Sec. \ref{sec:gr-direction} we repeat this parameter study for the force-free magnetic field configuration (see Sec. \ref{sec:setup2}), where we also vary the plasma-$\beta$.

\subsection{Harris current sheet}\label{sec:gr-harris}
For the Harris current sheet setup (Sec. \ref{sec:setup1}), the effect of shear flow on the resistive tearing mode was probed in Ref. \onlinecite{Li2010} using non-linear, incompressible MHD simulations by computing the reconnection rate for a selection of test cases. Here, we compute the linear growth rate using the compressible equations for various parameter combinations, initially to look at how the growth rate scales with resistivity, and later density. Assuming $\bfk = k_y\,\ey$, the shear ratio reduces to
\begin{equation}
    R_0 = \frac{a_B v_0}{a_v B_0}.
\end{equation}
Hence, we also verify that when the flow shear exceeds the magnetic shear, i.e. $R_0 > 1$, the tearing instability is fully stabilised, \cite{Chen1990} by varying the velocity parameters.

\subsubsection{Linear tearing of the Harris sheet}\label{sec:visualisation}
\begin{figure}[h]
    \centering
    \includegraphics[width=0.48\textwidth]{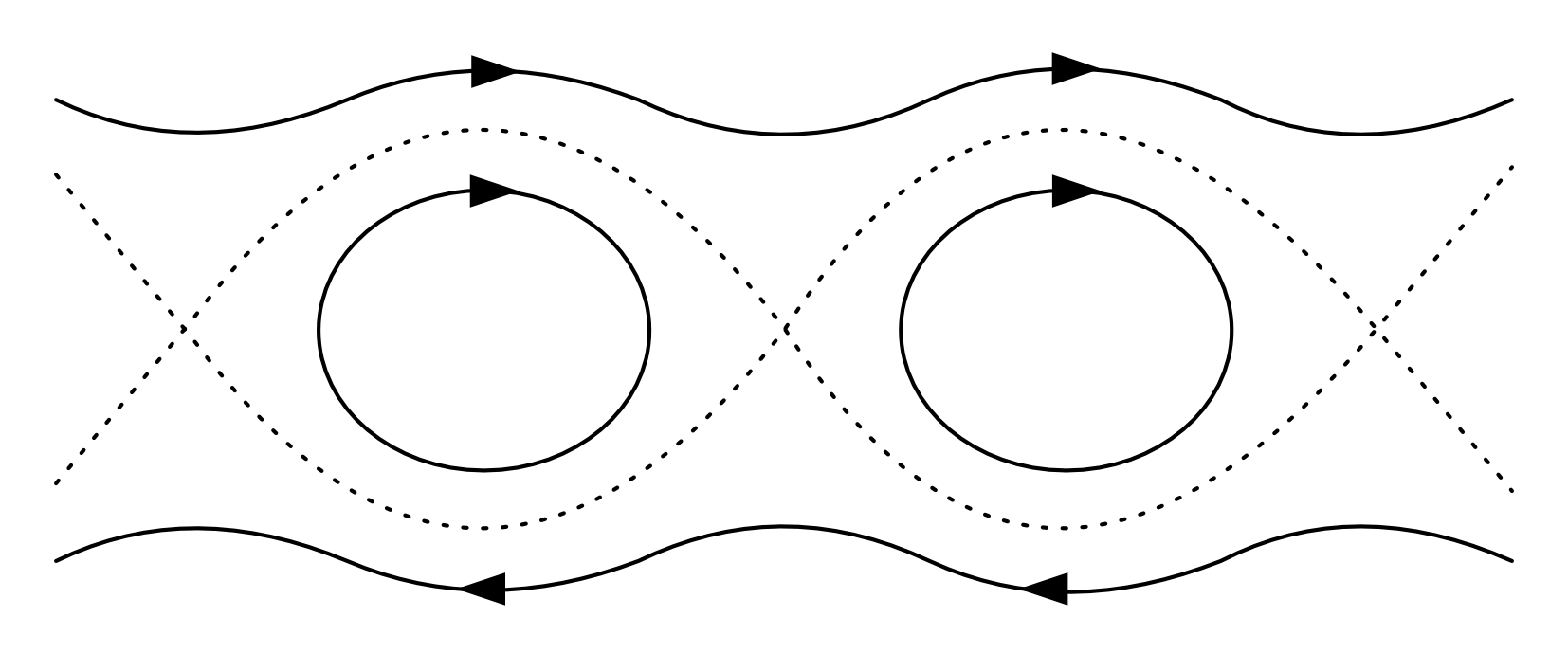}
    \caption{Schematic representation of magnetic field lines altered by the tearing instability.}
    \label{fig:schematic}
\end{figure}

To begin, we look at the influence of the resistive tearing mode on the magnetic field and flow. To do so, we consider the Harris sheet presented in Sec. \ref{sec:setup1}, both with and without the background flow, Eq. (\ref{eq:harris-flow}). In both cases, we set our parameters to $\bfk = 0.12\,\ey$, $\rho_0 = 1$, $B_\mathrm{c} = 1$, and $a_B = 1$, and use $v_\mathrm{c} = 0.25$ and $a_v = 0.75$ when equilibrium flow is included. This is solved on a symmetric, accumulated grid as described in App. \ref{app:grid} with parameters $p_1 = 0.2$, $p_2 = 0$, $p_3 = 0.01$, and $p_4 = 5$, resulting in $327$ grid points. In this configuration, the sheet is expected to tear up into magnetic islands, as represented schematically in Fig. \ref{fig:schematic}.

\begin{figure*}
    \centering
    \includegraphics[width=\textwidth]{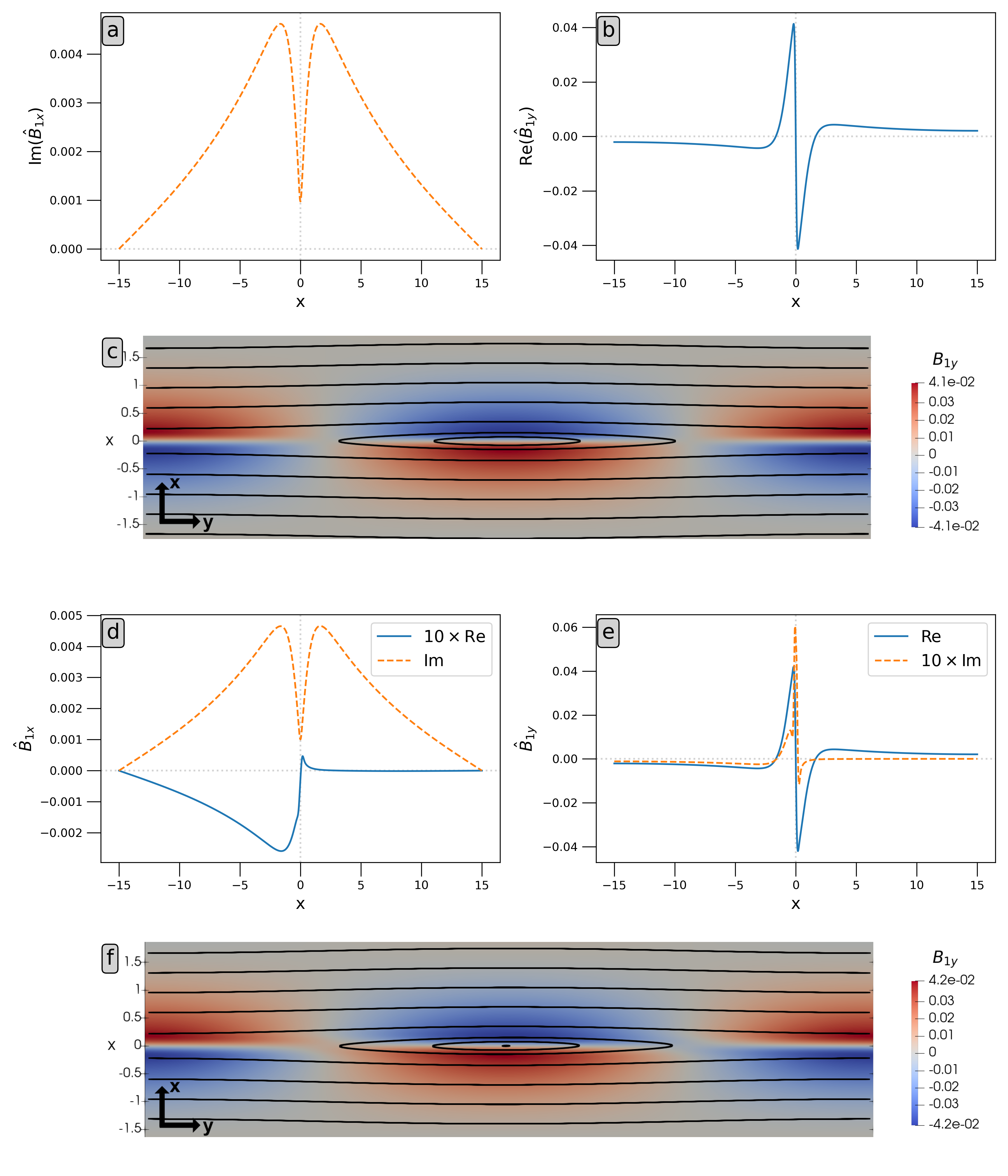}
    \caption{Magnetic field during linear Harris sheet tearing. No flow: (a) $\hat{B}_x$ perturbation amplitude; (b) $\hat{B}_y$ perturbation amplitude; (c) magnetic field lines and $B_{1y}$-visualisation (colour) for $\max(\rho_1) = 0.01\,\rho_0$. With flow: (d) $\hat{B}_x$ perturbation amplitude; (e) $\hat{B}_y$ perturbation amplitude; (f) magnetic field lines and $B_{1y}$-visualisation (colour) for $\max(\rho_1) = 0.01\,\rho_0$.}
    \label{fig:vis-B}
\end{figure*}

For the flowless case, Figs. \ref{fig:vis-B}(a,b) show \Legolas{}'s solutions for the $\hat{B}_x$ and $\hat{B}_y$ perturbation amplitudes from Eq. (\ref{eq:fourier}), respectively, where the arbitrary factor was chosen in such a way that the maximal density perturbation equals $1\%$, i.e. $\max(\rho_1) = 0.01\,\rho_0$, at $t = 0$. Similarly, Figs. \ref{fig:vis-B}(d,e) show the amplitudes in the case with background flow. In either case, $B_z$ is not perturbed. Note that in the flowless case one component is purely real and one is purely imaginary whilst the presence of the background shear flow makes both components fully complex.

Substituting these perturbation amplitudes in Eq. (\ref{eq:fourier}) for $t = 0$ and $y$ ranging from $-0.5\lambda$ to $0.5\lambda (\simeq 26.18)$, with $\lambda = 2\pi/|\bfk|$ the wavelength, results in the linear perturbation of the magnetic field, presented as field lines and with a colour map of the $B_{1y}$-component, in Figs. \ref{fig:vis-B}(c) and (f), for the case without and with flow, respectively. Note that since the amplitudes are normalised and can be multiplied with any complex factor, in principle, the (dimensionless) time $t$ presented here cannot be linked to a physical time without context. Further note that the range of $x$-coordinates was limited to focus on the behaviour near the sheet. As expected from the schematic representation in Fig. \ref{fig:schematic}, the visualisation in Fig. \ref{fig:vis-B}(c) reveals a magnetic island in the centre, with dipping in the previously straight magnetic field lines on either side. If the system were to evolve linearly, however, due to the sharp peaks in the $B_y$ perturbation amplitude on either side of the magnetic nullplane (see panel b), the magnetic field lines inside the island would dip progressively deeper inwards at the nullplane (because they cannot cross), until a magnetic field line meets itself again at the origin, dividing the island into two smaller islands (closed magnetic fieldlines) on either side of the nullplane. Since this is not observed in non-linear simulations, the time when this behaviour starts to develop in the linear solution marks an upper bound on the transition time from the linear to the non-linear regime. The inclusion of background shear flow does not dramatically alter this magnetic field structure, aside from introducing a slight shear deformation of the field lines across the nullplane. For the chosen parameters, this cannot be seen clearly at $t = 0$, shown in Fig. \ref{fig:vis-B}(f), but the effect becomes more apparent as the perturbation grows.

\begin{figure*}
    \centering
    \includegraphics[width=\textwidth]{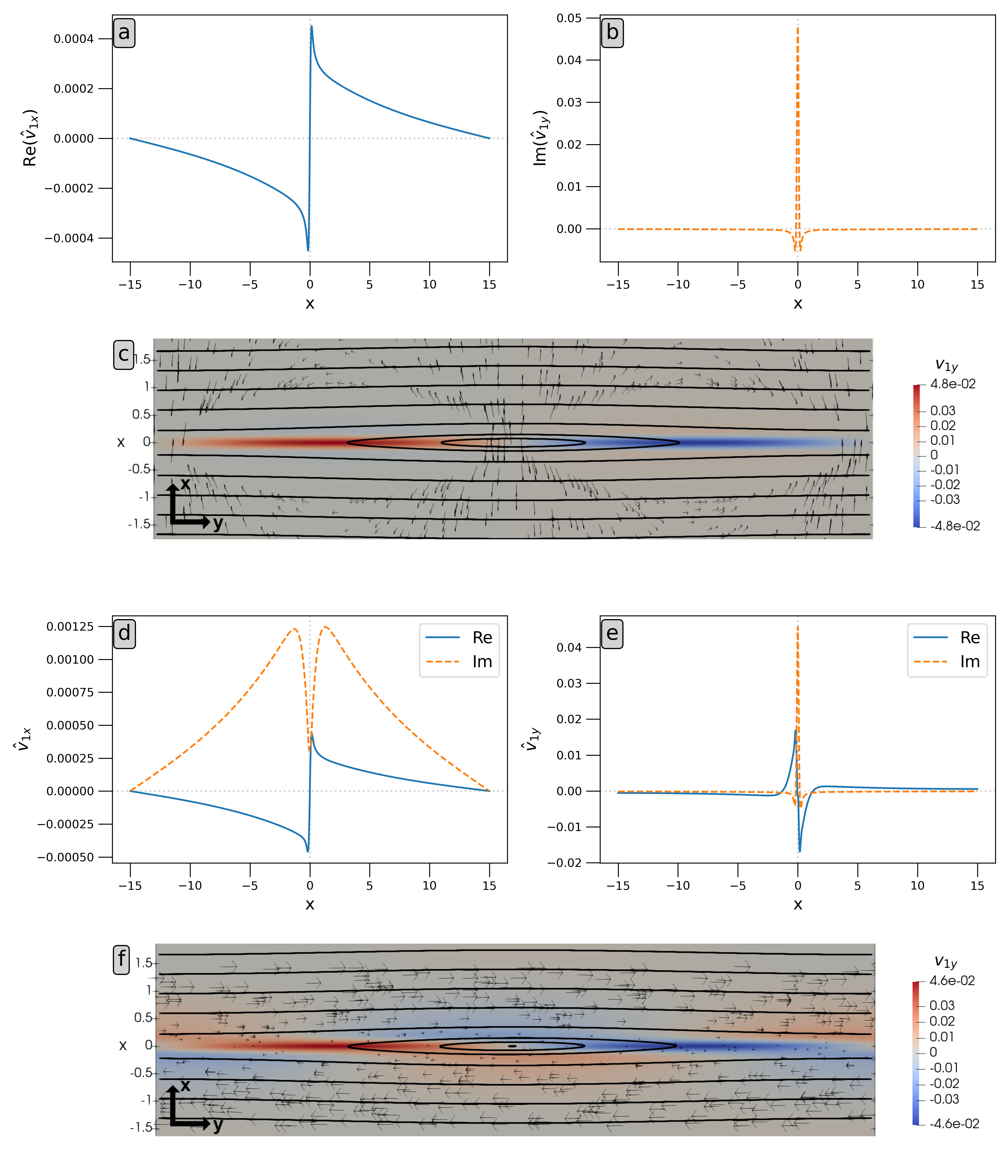}
    \caption{Flow during linear Harris sheet tearing. No flow: (a) $\hat{v}_x$ perturbation amplitude; (b) $\hat{v}_y$ perturbation amplitude; (c) magnetic field lines, stream vectors for $|\bfv| < 0.5\times 10^{-3}$, and $v_{1y}$-visualisation (colour) for $\max(\rho_1) = 0.01\,\rho_0$. With flow: (d) $\hat{v}_x$ perturbation amplitude; (e) $\hat{v}_y$ perturbation amplitude; (f) magnetic field lines, (fixed-length) stream vectors for $|\bfv| < 0.24$, and $v_{1y}$-visualisation (colour) for $\max(\rho_1) = 0.01\,\rho_0$.}
    \label{fig:vis-v}
\end{figure*}

Similarly, Fig. \ref{fig:vis-v} presents the $\hat{v}_x$ and $\hat{v}_y$ perturbation amplitudes from Eq. (\ref{eq:fourier}) for the flowless case in panels (a,b), and for the case with flow in panels (d,e). Again, the $z$-component, $v_z$, is not perturbed, and the flowless case has a purely real and purely imaginary $\bfv_1$-component. Identically to the magnetic field perturbation, the velocity perturbation amplitude also becomes complex when background flow is added. Contrary to the magnetic field perturbation amplitude though, where the newly introduced real/imaginary parts are an order of magnitude smaller than the original part, the real and imaginary parts of the velocity perturbation amplitudes are of a comparable order of magnitude.

Once more, Figs. \ref{fig:vis-v}(c,f) display a 2D visualisation of the magnetic field lines for the same $y$-interval and time as Figs. \ref{fig:vis-B}(c,f). Here, however, the panels are coloured by the magnitude of the $y$-component of the velocity perturbation, $v_{1y}$. In addition, the flow patterns are further highlighted with stream vectors up to a certain magnitude $|\bfv|$, to show variations in the regions of smaller speed. From the colour map in Fig. \ref{fig:vis-v}(c), it is immediately clear that the velocity in the flowless case is much higher at the magnetic nullplane than further away from it. In fact, the velocity indicates an inflow towards the centre of the magnetic island along the nullplane. Away from the sheet's centre, the stream vectors cross the magnetic field lines, moving outward from the island's centre, though with a much smaller speed than the inflow speed. The addition of background flow in Fig. \ref{fig:vis-v}(f) changes the picture significantly. Whilst the flow speed is still much higher in the centre of the sheet, the velocity now roughly follows the magnetic field at the island edges, resulting in a flow within the island. Again, the islands are too small at $t = 0$ to clearly distinguish the flow pattern inside, but it becomes apparent as the perturbation grows.

\subsubsection{Matching quantity and resistive layer halfwidth}\label{sec:harris-delta}
Before turning to the growth rate scaling, we evaluate the numerical matching quantity and resistive layer halfwidth, how they compare to analytic values, and how they are affected by introducing flow. Once again, we first consider the static Harris sheet with $\bfk = 0.5\,\ey$, $\rho_0 = 1$, $B_\mathrm{c} = 1$, and $a_B = 1$ ($\bfv_0 = 0$) in this section. For this static Harris sheet configuration, the analytic matching quantity is given exactly by\cite{Furth1963,Biskamp2000}
\begin{equation}\label{eq:Dp-harris}
    \Delta_\mathrm{A}' = \frac{2}{a_B} \left( \frac{1}{ka_B} - ka_B \right).
\end{equation}
Hence, for the chosen parameters we find $\Delta_\mathrm{A}' = 3$. Additionally, for constant-$\psi$ modes, the analytic resistive layer halfwidth is given by\cite{Biskamp2000}
\begin{equation}
    \delta_\mathrm{A} \simeq \eta^{2/5} (\Delta_\mathrm{A}')^{1/5} \left( k B_0' \right)^{-2/5},
\end{equation}
which evaluates to $\delta_\mathrm{A} \simeq 4.13\times 10^{-2}$ here, where $B_0'(x)$ is evaluated at the nullplane ($x=0$). Note that this satisfies the constant-$\psi$ condition $\delta_\mathrm{A} |\Delta_\mathrm{A}'| \simeq 0.124 < 1$.

Numerically (grid parameters $p_1 = 0.2$, $p_2 = 0$, $p_3 = 0.01$, $p_4 = 5$), as explained in Sec. \ref{sec:conventions}, we find $\Delta_\mathrm{N}' \simeq 2.56$ and $\delta_\mathrm{N} \simeq 3.5 \times 10^{-2}$ (note the use of $\simeq$ due to the $x$-discretisation in \Legolas{}). Again, the constant-$\psi$ condition is satisfied, $\delta_\mathrm{N} |\Delta_\mathrm{N}'| \simeq 0.090 < 1$. Hence, though the analytic and numerical values for $\Delta'$ and $\delta$ are not in perfect agreement, they are of the same order and reasonably close. Additionally, though the wavenumber $k=0.5$ is neither small nor large considering that the configuration is tearing unstable for $k \lesssim 1$, the constant-$\psi$ approximation holds, and thus we expect to recover a growth rate scaling proportional to $\eta^{3/5}$ in the static case.

\begin{figure*}
    \centering
    \includegraphics[width=\textwidth]{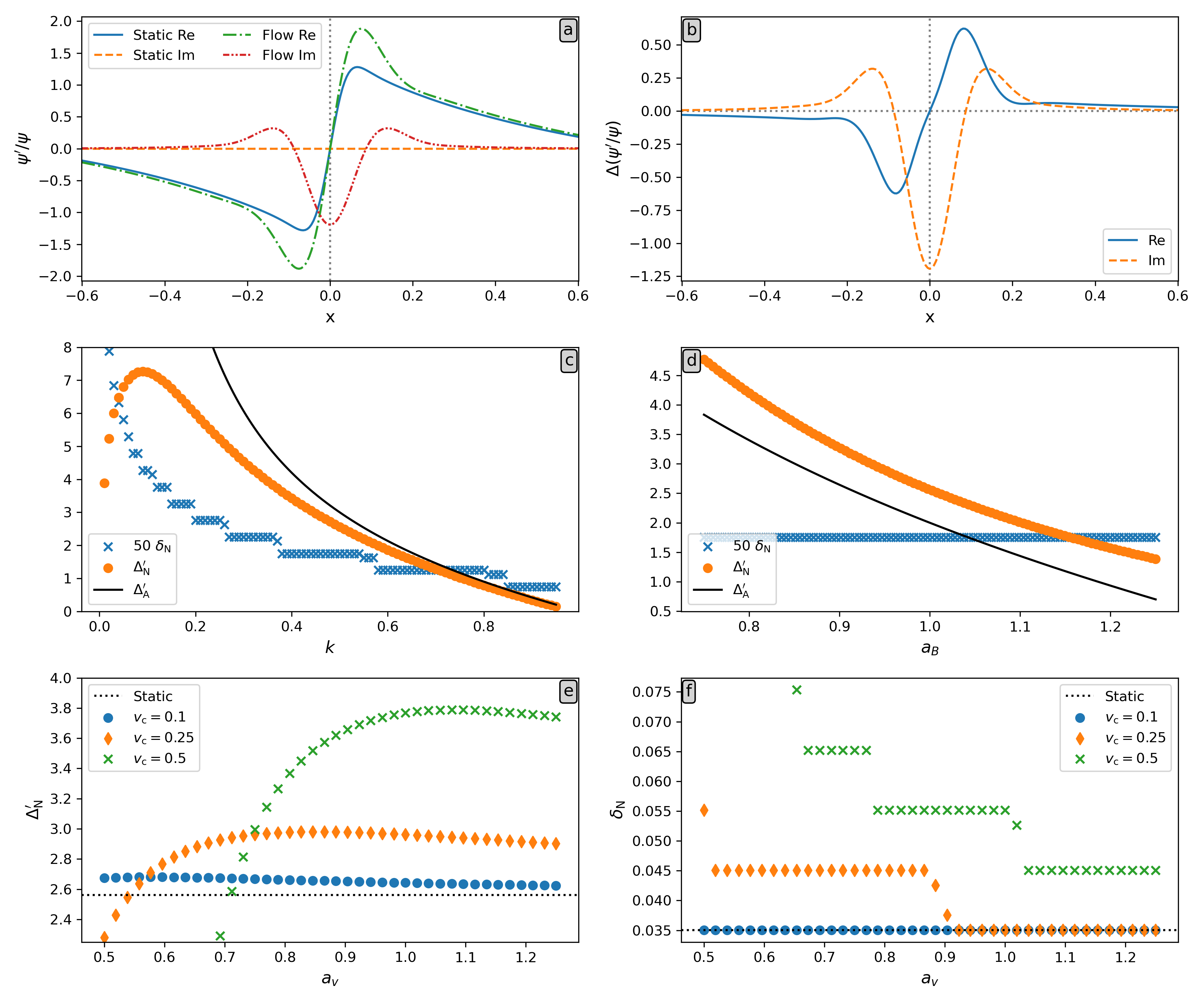}
    \caption{(a) $\psi'/\psi$ for $k = 0.5$ in a static and stationary Harris sheet. (b) Stationary minus static difference in $\psi'/\psi$ from panel (a). (c) Static $\Delta'$ and $\delta$ as a function of $\bfk = k_y\,\ey$. (d) Static $\Delta'$ and $\delta$ as a function of $a_B$. (e) $\Delta'_\mathrm{N}$ and (f) $\delta_\mathrm{N}$ for $k = 0.5$ in a Harris sheet with flow of various transition halfwidths.}
    \label{fig:psi-harris}
\end{figure*}

Analytically, Hofmann\cite{Hofmann1975} showed that if the velocity $\bfv_0$ is proportional to the Alfv\'en velocity (and thus the magnetic field $\bfb_0$) everywhere, the matching quantity is unaltered from the static case. For the Harris sheet with velocity profile Eq. (\ref{eq:harris-flow}), $\bfv_0 \propto \bfb_0$ is only true if $a_v = a_B$. However, as shown in Fig. \ref{fig:psi-harris}(a), the $\psi'/\psi$ ratio is not identical for the static and stationary ($v_\mathrm{c} = 0.5$, $a_v = 1$) case. This is further highlighted by the difference between both cases, shown in Fig. \ref{fig:psi-harris}(b). Consequently, neither is the numerical matching quantity. Now, we find a numerical matching quantity of $\Delta'_\mathrm{N} \simeq 3.77$ and a resistive layer halfwidth of $\delta_\mathrm{N} \simeq 5.5\times 10^{-2}$. Hence, both the numerical matching quantity and resistive layer halfwidth depend on the velocity, even if $\bfv_0 \propto \bfb_0$. However, $\delta_\mathrm{N}|\Delta'_\mathrm{N}| \simeq 0.21 < 1$ still holds. Since $\psi'/\psi$ is identical in our incompressible approximation, which also eliminates the Joule heating,\cite{DeJonghe2022} this deviation from Hofmann's result is presumably due to the use of a constant resistivity rather than a convectively perturbed resistivity.

The discrepancy between analytic and numerical $\Delta'$ and $\delta$, as well as the difference between the static and stationary Harris sheet, raises the question how the numerical matching quantity and resistive layer halfwidth depend on the various parameters. In Fig. \ref{fig:psi-harris}(c), both the analytic and numerical matching quantities are shown side by side as a function of $k$ for the static Harris sheet. For large wavenumbers, i.e. in the constant-$\psi$ regime, the analytic and numerical $\Delta'$ are in good agreement, but for small wavenumbers, the analytic matching quantity diverges to $+\infty$, whereas $\Delta'_\mathrm{N}$ is observed to decrease again. The maximal value of $\Delta'_\mathrm{N}$ in our set of discrete $k$-values was achieved for $k = 0.09$, with $\delta_\mathrm{N} |\Delta'_\mathrm{N}| \simeq 0.62$, indicating a strong numerical deviation from the analytic results in the traditional nonconstant-$\psi$ regime. Simultaneously, no maximum is observed in $\delta_\mathrm{N}$, which continues to increase as $k$ decreases. Note that the steplike behaviour of $\delta_\mathrm{N}$ is due to the discretisation of the $x$-coordinate in \Legolas{} (also in panel f). This deviation from analytic results in the nonconstant-$\psi$ regime is not surprising. As $k$ decreases, $\psi(0)$ approaches zero and $\Delta'_\mathrm{N}$ approaches an $x^{-1}$ scaling. Due to the definition of $\Delta'_\mathrm{N}$, the evaluation of $\psi'/\psi$ occurs near the edge of the resistive layer, and thus $\Delta'_\mathrm{N} \sim \delta_\mathrm{N}^{-1}$. Hence, $\Delta'_\mathrm{N}$ decreases as $\delta_\mathrm{N}$ increases for $k \rightarrow 0$. Consequently, $\Delta'_\mathrm{N}$ is not a proper numerical equivalent of $\Delta'_\mathrm{A}$ in the nonconstant-$\psi$ regime, and thus $\delta_\mathrm{N} |\Delta'_\mathrm{N}|$ cannot be used to distinguish between constant-$\psi$ and nonconstant-$\psi$ modes.

Similarly to Fig. \ref{fig:psi-harris}(c), Fig. \ref{fig:psi-harris}(d) presents the dependence of $\Delta'_\mathrm{N}$ and $\delta_\mathrm{N}$ on the magnetic reversal halfwidth $a_B$. Here, the numerical matching quantity appears to follow the analytic scaling reasonably well, though it is consistently larger than $\Delta'_\mathrm{A}$ for our choice of $k = 0.5$. The numerical resistive layer halfwidth $\delta_\mathrm{N}$, on the other hand, is observed to be constant as a function of $a_B$, which is surprising considering that Furth \etal\cite{Furth1963} derived an $a_B^{-1}$ dependence for the width of the region of discontinuity.

Finally, again adding the velocity profile Eq. (\ref{eq:harris-flow}), Figs. \ref{fig:psi-harris}(e,f) show the dependence of $\Delta'_\mathrm{N}$ and $\delta_\mathrm{N}$ on the transition halfwidth $a_v$ of the velocity profile for various flow speeds $v_\mathrm{c}$. As expected, the effect of flow on the matching quantity increases with the flow speed $v_\mathrm{c}$. For sufficiently high $v_\mathrm{c}$ and small $a_v$, the numerical matching quantity is observed to increase initially with $a_v$, whereas the resistive layer halfwidth decreases. As $a_v$ increases further, $\Delta'_\mathrm{N}$ reaches a maximum and decreases again steadily. For the resistive layer halfwidth, we observe a monotone decrease with $a_v$, which is more pronounced for larger $v_\mathrm{c}$.

\subsubsection{Growth rate scaling with resistivity}
In the seminal work by Furth, Killeen, and Rosenbluth \cite{Furth1963} the authors derive power laws for the scaling of the incompressible tearing growth rate as a function of the resistivity $\eta$, for small values of $\eta$. Here, we introduce compressibility and consider a wide range of resistivity values. However, contrary to their work, we assume a constant resistivity without convective perturbations, and include Joule heating. From their derivations they conclude that the growth rate scales as a power law in $\eta$, $\text{Im}(\omega) \sim \eta^p$, and that $p$ depends on whether or not $\psi$ is approximately constant across the magnetic nullplane. This $\psi$-classification remains equally important when shear flow is added.\cite{Chen1990}

\begin{figure*}
    \centering
    \includegraphics[width=\textwidth]{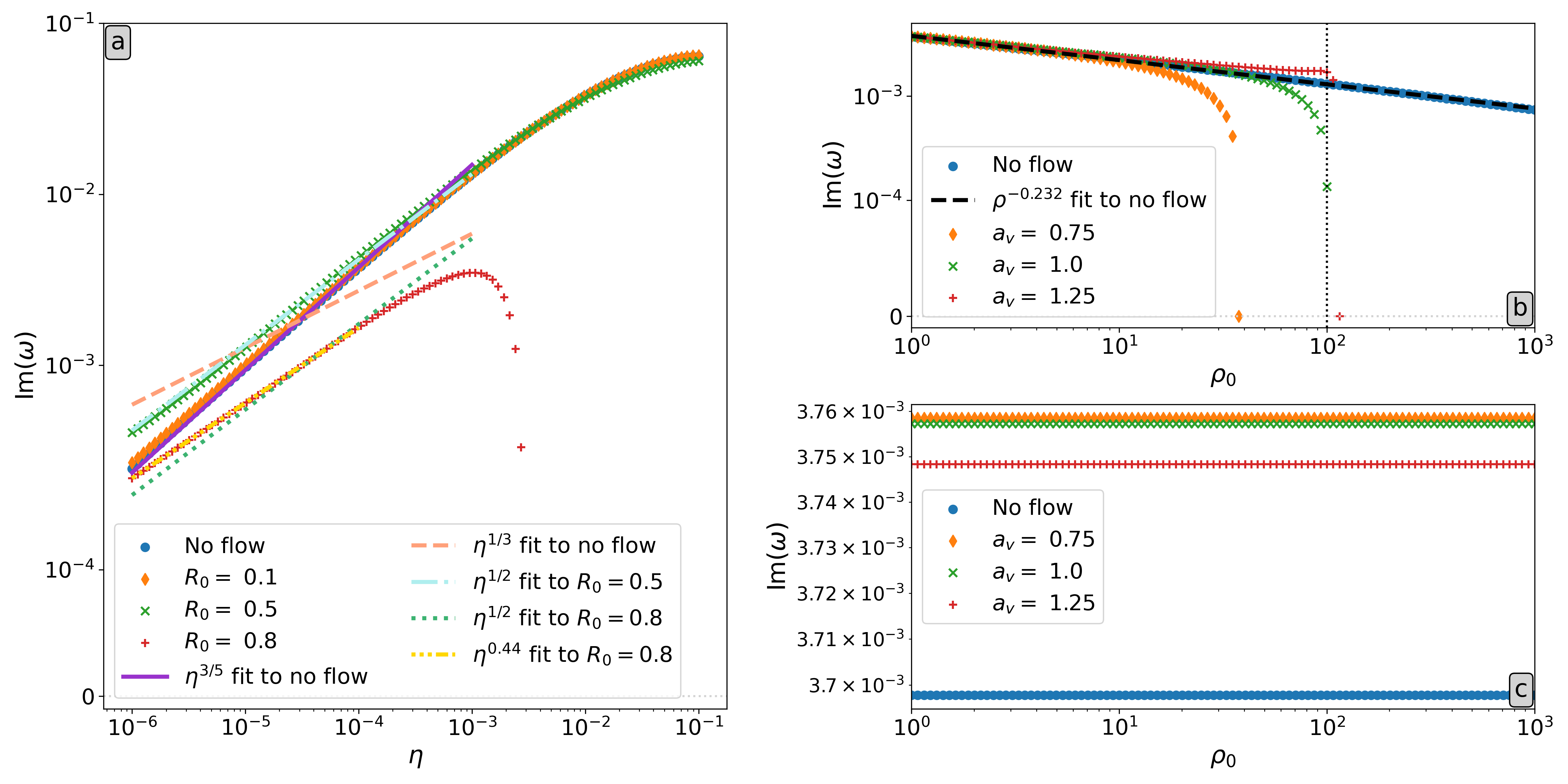}
    \caption{Resistive tearing growth rate of a Harris sheet, Eq. (\ref{eq:harris-B}), as a function of (a) the resistivity $\eta$; (b) the density $\rho_0$; for parameters $\bfk = 0.5\,\ey$, $\rho_0 = 1$, $B_\mathrm{c} = 1$, $a_B = 1$, and (a) $a_v = 1$; (b) $v_\mathrm{c} = 0.1$, if flow was included. In (b), the density where the maximal velocity of the equilibrium configuration coincides with the Alfv\'en speed is indicated with a dotted line. (c) Growth rate scaling with $\rho_0$ for constant Alfv\'en speed ($B_\mathrm{c} = \sqrt{\rho_0}$) and $p_\mathrm{c} = 10^6$.}
    \label{fig:harris-rhoeta}
\end{figure*}

In the last section, we have established that the tearing mode of the Harris sheet from Sec. \ref{sec:setup1} falls into the constant-$\psi$ regime for the parameters $\bfk = 0.5\,\ey$, $\rho_0 = 1$, $B_\mathrm{c} = 1$, $a_B = 1$, $a_v = 1$, and $v_\mathrm{c}$ sufficiently small. Now, we vary the resistivity $\eta$ to compare to the literature's corresponding scaling laws. In Fig. \ref{fig:harris-rhoeta}(a), the flowless case is compared to the case with flow profile Eq. (\ref{eq:harris-flow}) for a selection of $R_0$-values, which are expected to scale differently based on the analytic power laws presented in Ref. \onlinecite{Chen1990}. (Note that the no flow case is not clearly visible since it almost coincides with the $R_0 = 0.1$ case.) All \Legolas{} runs were performed using grid parameters $p_1 = 0.2$, $p_2 = 0$, $p_3 = 0.01$, and $p_4 = 5$ ($327$ grid points). The desired value of $R_0$ was obtained by setting $v_\mathrm{c} = a_v B_\mathrm{c} R_0 / a_B$.

Since the analytic scaling laws are derived under the assumption that there is a clear temporal separation between the resistive diffusion time $\tau_\mathrm{R} = \eta^{-1}$ (in our dimensionless units) and the Alfv\'en time $\tau_\mathrm{A} = a_B \sqrt{\rho_0}\ B_\mathrm{c}^{-1}$,\cite{Furth1963,Betar2022} comparison to these power laws is only meaningful for $\eta < 10^{-2}$ in this case. Therefore, to be on the safe side, all fitted power laws from Table \ref{tab:scaling-laws} were limited to $\eta < 10^{-3}$ in Fig. \ref{fig:harris-rhoeta}(a). In accordance with the work of Chen and Morrison,\cite{Chen1990} the scaling of the static Harris sheet's growth rate is found to be close to $\eta^{3/5}$, as expected in the constant-$\psi$ regime. For $R_0 = 0.1 \ll 1$, the growth rate is slightly larger than the static case, but the same scaling law seems to hold. As $R_0$ increases to $R_0 = 0.5$, the scaling law changes to $\eta^{1/2}$, which is also in line with the literature for $R_0 \lesssim 1$. However, for $R_0 = 0.8$, none of the analytic power laws are a good fit. At small $\eta$, this case is observed to scale as $\sim\eta^{0.44}$. Note that this power law lies between the constant-$\psi$ $\eta^{1/2}$ and nonconstant-$\psi$ $\eta^{1/3}$ scaling laws, though the product $\delta_\mathrm{N}|\Delta'_\mathrm{N}|$ remains significantly smaller than $1$, once again confirming it is not a good metric to classify $\psi$. In addition, the growth rate deviates strongly from a power law well before $\eta$ reaches $10^{-2}$.

As expected, aside from the growth rate dropoff in the $R_0 = 0.8$ case, the other cases also deviate from the power laws as $\eta$ crosses the $10^{-2}$ threshold. Note that growth rates above $10^{-2}$ should be interpreted with care, since diffusion of the equilibrium was neglected and might affect the dynamics significantly.

\subsubsection{Density variation}
To demonstrate that the growth rate does not care about the specific density value, but only about the Alfv\'en speed, first consider the static Harris sheet with $\bfk = 0.5\,\ey$, $B_\mathrm{c} = 1$, $a_B = 1$, and varying density (grid parameters: $p_1 = 0.2$, $p_2 = 0$, $p_3 = 0.01$, $p_4 = 5$). This is shown in Fig. \ref{fig:harris-rhoeta}(b), where a fit to the data shows that the growth rate scales approximately as $\rho_0^{-0.232}$. Comparing this to the analytic (incompressible) growth rate scaling $\rho_0^{-1/5}$ (see e.g. Ref. \onlinecite{Furth1963}, Eq. ($54$)), whilst reasonably close, the deviation is significant. This may be due to compressibility, especially considering that the plasma-$\beta$ goes to $\infty$ at the nullplane in this configuration.

Since both the Alfv\'en speed $c_\mathrm{A}$ and sound speed $c_\mathrm{s}$ are proportional to $\rho^{-1/2}$ (due to our choice of $T_0 \propto \rho_0^{-1}$ profile), the question becomes how the growth rate scales with either speed. To answer this, we now consider a second static Harris sheet with the same parameters as before, except we set $B_\mathrm{c} = \sqrt{\rho_0}$ and $p_\mathrm{c} = 10^6$. These choices ensure that the Alfv\'en speed is constant when varying the density whilst $T_0 \simeq p_\mathrm{c} / \rho_0$, and thus $c_\mathrm{s} \sim \rho_0^{-1/2}$. In this case we find a constant growth rate, shown in Fig. \ref{fig:harris-rhoeta}(c). Hence, we conclude that the growth rate does not depend on the specific density or sound speed, but scales with the Alfv\'en speed.

The effect of the inclusion of a non-zero flow of fixed size $v_\mathrm{c} = 0.1$ (and various transition halfwidths) depends on the case. In the latter case, the $v_\mathrm{c}/c_\mathrm{A}$-ratio is constant, and the growth rate modification does not vary with $\rho_0$. In the first case however, the flow modifies the tearing growth rate variation as a function of the density $\rho_0$, or better, because the ratio $v_\mathrm{c}/c_\mathrm{A}$ changes. In fact, the flow introduces a critical density above which the tearing mode is fully damped as $v_\mathrm{c}/c_\mathrm{A} \rightarrow 1$. This is highlighted in Fig. \ref{fig:harris-rhoeta}(b) by the dotted line representing where $v_\mathrm{c} = c_\mathrm{A}$. However, from the figure it is clear that the tearing mode is not always damped exactly when $v_\mathrm{c}$ reaches $c_\mathrm{A}$, but depends on the flow transition halfwidth $a_v$. Therefore, the Alfv\'en speed does not necessarily act as a transition value in the equilibrium speed with respect to tearing suppression. Though the tearing instability may vanish at a density lower than where the equilibrium speed equals the Alfv\'en speed, it appears the Alfv\'en speed still imposes an upper limit on the density above which the tearing mode vanishes, from the sharp dropoff there in the $a_v = 1.25$ case.

\subsubsection{Velocity variation}
\begin{figure*}
    \centering
    \includegraphics[width=\textwidth]{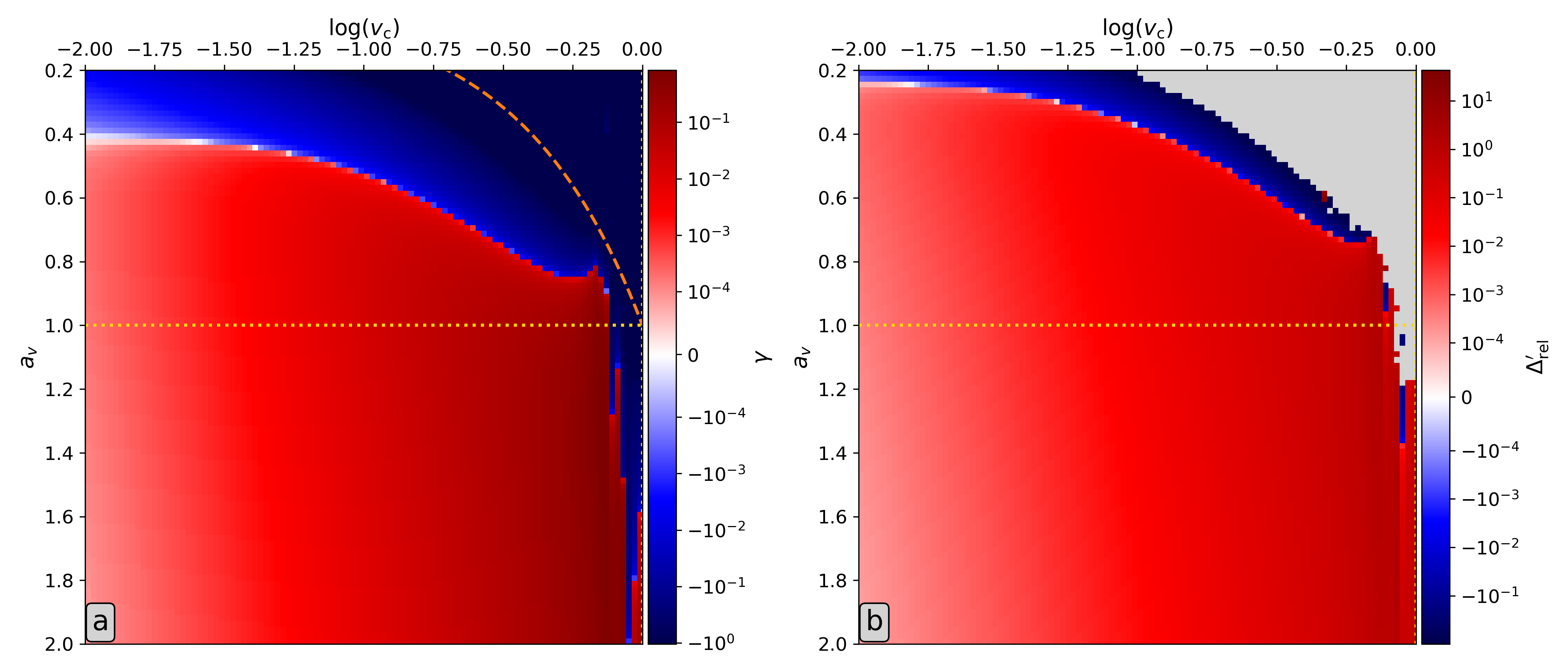}
    \caption{(a) Relative tearing growth rate $\gamma$ and (b) relative numerical matching quantity $\Delta'_\mathrm{rel}$ with respect to the static case for combinations of the maximal speed $v_\mathrm{c}$ and flow transition halfwidth $a_v$. The yellow dotted line indicates the magnetic field transition halfwidth $a_B$ and the orange dashed line represents $R_0 = 1$.}
    \label{fig:harris-grid-v02av}
\end{figure*}

Since both parameters of the velocity profile (the maximal speed $v_\mathrm{c}$ and halfwidth $a_v$) appear to play an important role, we vary both parameters simultaneously to identify the regions of stabilisation and further destabilisation of the tearing instability. Here, the maximal speed $v_\mathrm{c}$ is kept sub-Alfv\'enic ($v_\mathrm{c} < 1$) because the flow-induced KHI dominates in the super-Alfv\'enic regime.\cite{Hofmann1975} The result is shown in Fig. \ref{fig:harris-grid-v02av}(a) for fixed parameters $\bfk = 0.5\,\ey$, $\rho_0 = 1$, $B_0 = 1$, and $a_B = 1$, and was obtained with the shift-invert method on an accumulated grid with parameters $p_1 = 0.2$, $p_2 = 0$, $p_3 = 0.01$, and $p_4 = 5$ ($327$ grid points). Since this method requires an initial guess, the first guess was obtained from a run with the \texttt{QR-cholesky} solver for $v_\mathrm{c} = 10^{-2}$ and $a_v = 1$, which was used to compute the growth rate for all values of $a_v$ and $v_\mathrm{c} = 10^{-2}$ with the shift-invert method. This array of growth rates was subsequently used as the initial guesses for the next value of $v_\mathrm{c}$, and so on. Visually speaking, each value (except those in the left column) in Fig. \ref{fig:harris-grid-v02av}(a) was computed via shift-invert by providing the value to its immediate left as the initial guess. Note that whilst the tearing mode is fully damped in the top right corner of panel (a), the system is still unstable because the KHI appears here before $v_\mathrm{c}$ reaches the Alfv\'en speed. This was again checked with the \texttt{QR-cholesky} solver in this region of the parameter space. (For the few parameter combinations where the shift-invert method failed to converge, the growth rate was calculated using the \texttt{QR-cholesky} solver.) Since the analytic scalings laws typically feature a scaling with the matching quantity, Fig. \ref{fig:harris-grid-v02av}(b) shows the relative numerical matching quantity $\Delta'_\mathrm{rel}$ for the parameter combinations where the tearing growth rate exceeds $10^{-4}$, and grey elsewhere.

In this figure, a few things stand out. First of all, the resistive tearing instability is clearly stabilised at the $R_0 = 1$ (orange dashed) line for all considered parameter combinations, but is already fully suppressed before this line is reached. Since the analytic growth rate scales with $\Delta'_\mathrm{A}$\cite{Chen1990} and $\Delta'_\mathrm{A}$ is only identical to the static case if $\bfv_0 \propto \bfb_0$,\cite{Hofmann1975} the instability is only expected to vanish exactly at $R_0 = 1$ if $a_v = a_B$, so this is not surprising. However, even for $a_v = a_B$ (yellow dotted line), the instability is already fully suppressed before $R_0 = 1$ is reached. This may be due to the earlier observation (see Sec. \ref{sec:harris-delta}) that $\Delta'_\mathrm{N}$ does differ between static and stationary Harris sheets here, even if $\bfv_0 \propto \bfb_0$.

Secondly, contrary to the non-linear observation in Ref. \onlinecite{Li2010} that there exists a single critical $a_v$ value $\sim 0.35$ where the transition from stabilising to destabilising occurs, we here observe that this critical $a_v$ depends on the maximal speed $v_\mathrm{c}$. Furthermore, their critical value lies in our stabilising region of the parameter regime across all velocities. Note though that the simulations in Ref. \onlinecite{Li2010} are incompressible, include a non-zero viscosity, and lack Joule heating, any of which may affect this result.

Finally, comparing Figs. \ref{fig:harris-grid-v02av}(a) and (b), there is a discrepancy between the variations in the growth rate and the numerical matching quantity. At smaller $a_v$, the growth rate is slightly stabilised whilst the matching quantity is slightly increased. Therefore, since we only varied the flow parameters, the influence of shear flow on the growth rate is not fully encapsulated in the numerical matching quantity, and the growth rate does not simply scale with $\Delta'_\mathrm{N}$.

\subsection{Force-free magnetic field}\label{sec:gr-direction}
Now, we turn to the setup from Sec. \ref{sec:setup2}, where a magnetic field of fixed magnitude varies its direction periodically throughout the plasma slab. In this particular case, the shear ratio becomes
\begin{equation}
    R_0 = \left| \frac{G'(0)}{F'(0)} \right| = \frac{v_\mathrm{c} \sqrt{\rho_\mathrm{c}}}{\alpha}.
\end{equation}
Whilst there are three parameters in this expression, our parametric study will only focus on the variation of the equilibrium density $\rho_\mathrm{c}$ and velocity coefficient $v_\mathrm{c}$, for reasons which will become clear after a demonstration of the role of $\alpha$.

\subsubsection{Multiple tearing modes}
\begin{figure*}
    \centering
    \includegraphics[width=0.93\textwidth]{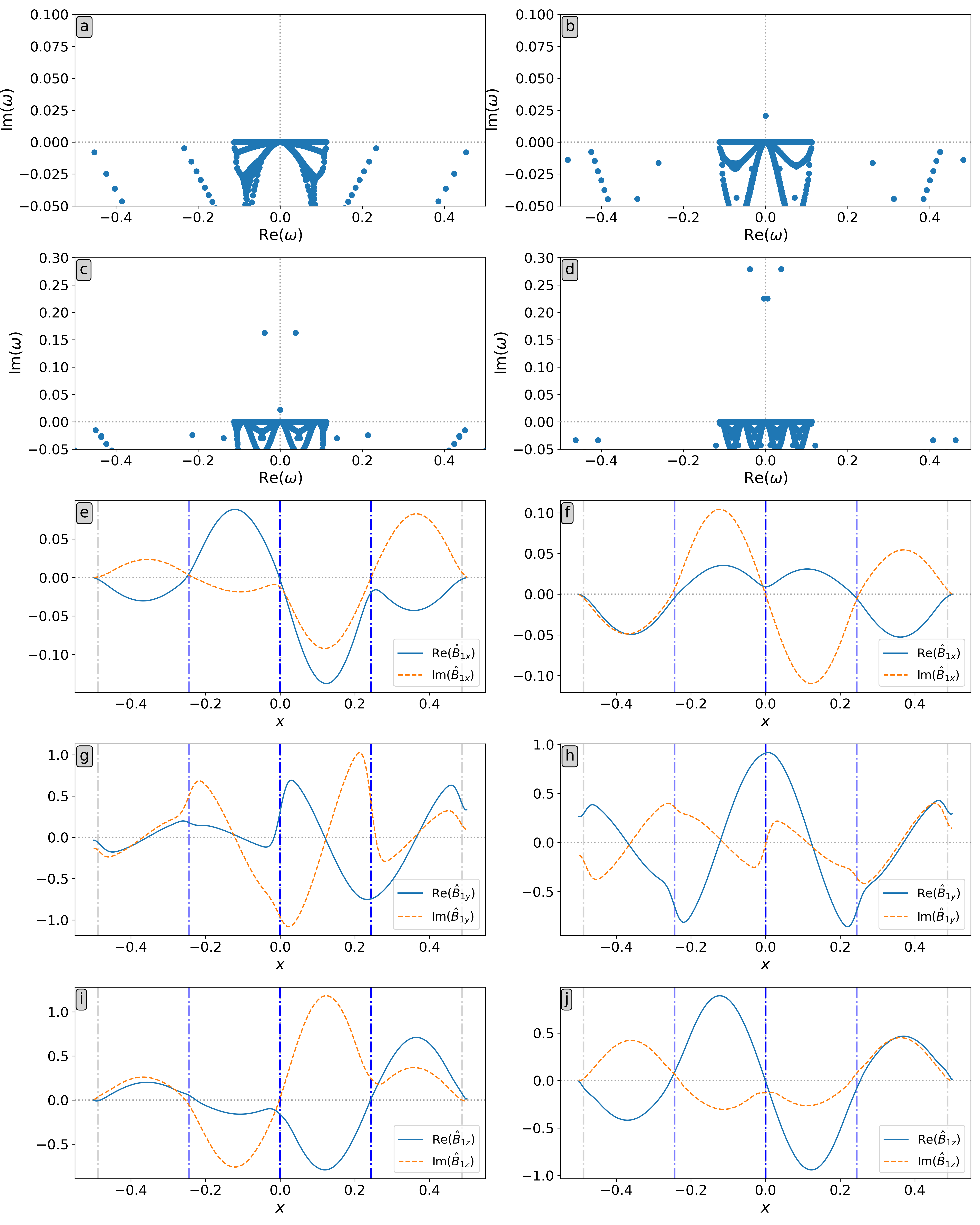}
    \caption{(a-d) Parts of the spectra of a plasma slab with force-free magnetic field, Eqs. (\ref{eq:equil-rotB}), for $\rho_\mathrm{c} = 1$, $\beta_0 = 0.15$, $v_\mathrm{c} = 0.15$, and $\bfk = 1.5\,\ey$. The angular parameter $\alpha$ in the magnetic field profile determines the number of magnetic nullplanes and unstable modes, and takes a different value in each panel: (a) $\alpha = \pi/2$, (b) $\alpha = 4.73884$, (c) $\alpha = 5\pi/2$, and (d) $\alpha = 4.1\pi$. (e-j) $\hat{\bfb}_1$-eigenfunctions of panel (d)'s (left: e, g, i) dominant, forward-travelling tearing mode; (right: f, h, j) less unstable, forward-travelling tearing mode. Magnetic nullplanes are indicated by dash-dotted lines, with blue lines marking where tearing occurs (lighter blue indicates that multiplication with a complex factor is required to highlight tearing behaviour there).}
    \label{fig:multitearing}
\end{figure*}

\begin{figure*}
    \centering
    \includegraphics[width=\textwidth]{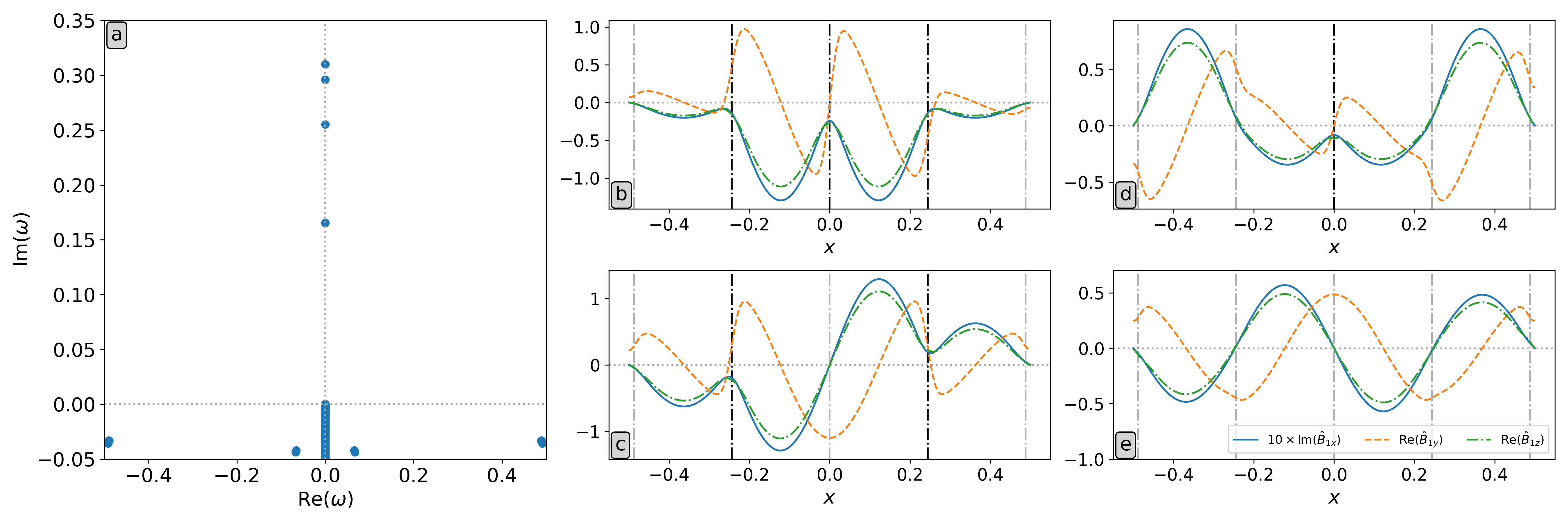}
    \caption{(a) Part of the spectrum of a plasma slab with force-free magnetic field, Eqs. (\ref{eq:equil-rotB}), for $\rho_\mathrm{c} = 1$, $\beta_0 = 0.15$, $\alpha = 4.1\pi$, and $\bfk = 1.5\,\ey$, without background flow ($v_\mathrm{c} = 0$). (b-e) $\hat{\bfb}_1$-eigenfunctions for (b) $\omega \simeq 0.3100\,\im$; (c) $\omega \simeq 0.2961\,\im$; (d) $\omega \simeq 0.2553\,\im$; (e) $\omega \simeq 0.1654\,\im$. Magnetic nullplanes are indicated by dash-dotted lines, with darker lines marking where tearing occurs in that panel.}
    \label{fig:multitearing-noflow}
\end{figure*}

Since the parameter $\alpha$ regulates how fast the magnetic field's direction varies along $x$, it also determines how many magnetic nullplanes the system has in a certain $x$-interval for a given wave vector. Consequently, the number of tearing modes supported by the configuration depends on $\alpha$. Additionally, if the equilibrium velocity is described by an odd function, like the linear profile in Eqs. (\ref{eq:equil-rotB}), the spectrum is symmetric with respect to the imaginary axis. This is illustrated in Figs. \ref{fig:multitearing}(a-d), where we varied the parameter $\alpha$ for parameters $\rho_\mathrm{c} = 1$, $\beta_0 = 0.15$, $v_\mathrm{c} = 0.15$, and $\bfk = 1.5\,\ey$ at $251$ grid points. For (a) $\alpha = \pi/2$, the magnetic shear is insufficient to induce a tearing instability. Increasing $\alpha$ without introducing an additional nullplane results in one non-propagating tearing mode (i.e. purely imaginary), visualised in (b) for $\alpha = 4.73884$ (slightly more than $3\pi/2$). In the presence of three nullplanes for $\alpha = 5\pi/2$, (c) shows a pair of forward-backward propagating instabilities and one non-propagating one. Finally, (d) contains only two pairs of forward-backward propagating tearing pairs for $\alpha = 4.1\pi$, despite the presence of $5$ nullplanes in the domain. If the equilibrium flow is removed, all tearing modes become non-propagating, i.e. purely imaginary, as demonstrated in Fig. \ref{fig:multitearing-noflow}(a) for the case with $\alpha = 4.1\pi$.

For the flowless case in Fig. \ref{fig:multitearing-noflow}(a), the magnetic field perturbation amplitudes of the unstable modes are shown in panels (b) through (e). All positions of magnetic nullplanes are marked with a dash-dotted line. At the darker-coloured nullplanes we observe a dip in $\hat{B}_{1x}$ and a sharp transition in $\hat{B}_{1y}$, similar to Figs. \ref{fig:vis-B}(a,b), indicative of tearing at this nullplane. Note that the dominant instability features tearing behaviour at the 3 central nullplanes, whereas the secondary instability only tears up the central nullplane whilst the tertiary mode is tearing up the same nullplanes as the dominant instability except for the central nullplane. Surprisingly, the least unstable mode does not show any signs of tearing. Additionally, the outer nullplanes do not show strong signs of tearing, presumably due to their proximity to the perfectly conducting boundaries, which exert a stabilising influence.

With the addition of flow, however, the situation changes. Now, the spectrum in Fig. \ref{fig:multitearing}(d) no longer has a single dominant instability, but a dominant pair and less unstable pair of forward-backward propagating instabilities. Furthermore, their perturbation amplitudes are now fully complex. For the dominant, forward-propagating ($\text{Re}(\omega) > 0$) instability, the $\hat{\bfb}_{1}$ perturbation amplitudes are shown in Figs. \ref{fig:multitearing}(e,g,i), and similarly, those of the less unstable, forward-propagating instability are shown in Figs. \ref{fig:multitearing}(f,h,j). The backward-propagating counterparts are not shown, but are the mirror image with respect to $x=0$ of the forward-propagating perturbation amplitudes. The magnetic nullplanes are again indicated by dash-dotted lines.

Contrary to the static cases, all four modes appear to tear all three central nullplanes, coloured in blue, though lighter blue lines indicate multiplication with a complex factor is needed to highlight the tearing behaviour. Additionally, the dominant, forward-propagating mode's largest amplitudes variations (i.e. strongest tearing) occur at the positive intermediate and central nullplanes, with smaller variations at the negative intermediate nullplane. The less unstable, forward-propagating mode, on the other hand, has its largest amplitudes variations at the central nullplane, with smaller variations at both intermediate nullplanes. Hence, by introducing flow the tearing behaviour of all modes is altered significantly.

From now on, the value of $\alpha$ is set to $\alpha = 4.73884$ (i.e. the value used in Fig. \ref{fig:multitearing}(b) and Ref. \onlinecite{Goedbloed2019}). This ensures that the magnetic field makes between one-half and a full rotation in the considered domain, $x\in [-0.5,0.5]$, resulting in a single nullplane at $x=0$ and a single tearing mode. Since there is only one nullplane, no multitearing occurs.\cite{Furth1973, Pritchett1980} A detailed look at multitearing is beyond the scope of this paper.

\subsubsection{Matching quantity and resistive layer halfwidth}
Like in the previous case of the Harris sheet, we evaluate the matching quantity and resistive layer halfwidth for the force-free magnetic field configuration before moving on to the scaling of the growth rate (for now, we set aside our conclusion from that section regarding $\Delta'_\mathrm{N}$). Again, we choose the parameters $\bfk = 1.5\,\ey$, $\rho_\mathrm{c} = 1$, $\alpha = 4.73884$, and $\beta_0 = 0.15$. When velocity is included, we set $v_\mathrm{c} = 0.25$. All runs were performed at $251$ grid points. Numerically, the static case yields $\Delta'_\mathrm{N} \simeq 5.67$ and $\delta_\mathrm{N} \simeq 1.4\times 10^{-2}$, whereas the stationary case yields $\Delta'_\mathrm{N} \simeq 5.84$ and $\delta_\mathrm{N} \simeq 1.4\times 10^{-2}$. Hence, the constant-$\psi$ condition is satisfied in both cases, with $\delta_\mathrm{N} |\Delta'_\mathrm{N}| \simeq 0.079 < 1$ and $\delta_\mathrm{N} |\Delta'_\mathrm{N}| \simeq 0.082 < 1$ for the static and stationary case, respectively. Therefore, we again expect a growth rate scaling proportional to $\eta^{3/5}$ in the next section.

\begin{figure*}
    \centering
    \includegraphics[width=\textwidth]{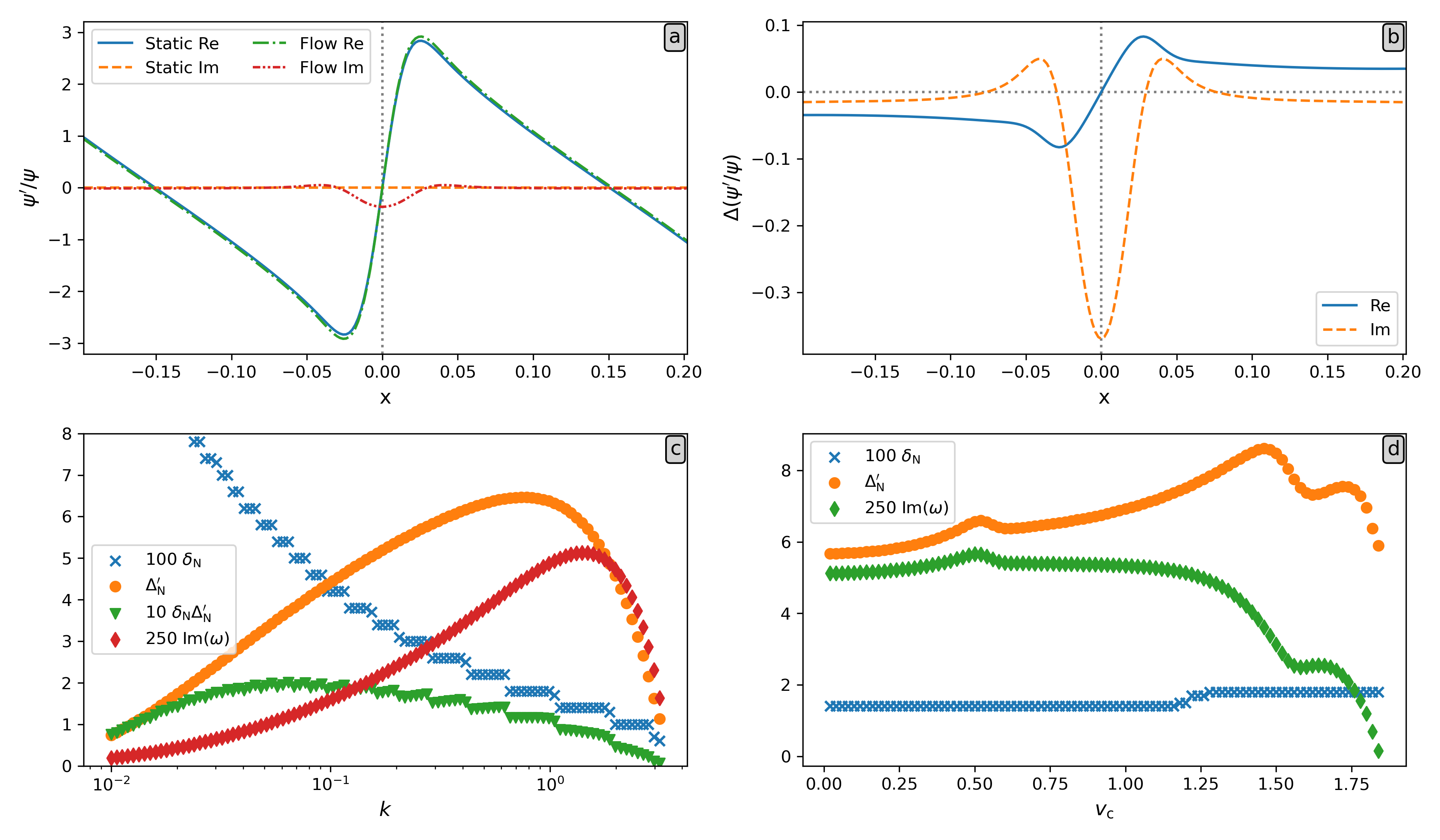}
    \caption{(a) $\psi'/\psi$ for $k = 1.5$ in the static and stationary force-free $\bfb_0$-configuration. (b) Stationary minus static difference in $\psi'/\psi$ from panel (a). (c) Static $\delta_\mathrm{N}$, $\Delta'_\mathrm{N}$, and tearing growth rate as a function of $\bfk = k_y\,\ey$. (d) Stationary $\delta_\mathrm{N}$, $\Delta'_\mathrm{N}$, and tearing growth rate as a function of $v_\mathrm{c}$.}
    \label{fig:psi-rotation}
\end{figure*}

Again, we look at the influence of the flow profile on the $\psi'/\psi$-ratio. In Fig. \ref{fig:psi-rotation}(a), the $\psi'/\psi$-ratio is shown for the static and stationary case. The difference between both cases is shown in Fig. \ref{fig:psi-rotation}(b). Clearly, the $\psi'/\psi$-ratio does not change wildly with the addition of this flow profile. However, the difference between both cases is not negligible either, explaining the difference in numerical matching quantity calculated above.

Now the question arises how strongly the numerical matching quantity is impacted by a variation in wavenumber or speed, and whether the growth rate scales with $\Delta'_\mathrm{N}$. In Fig. \ref{fig:psi-rotation}(c), the numerical matching quantity and resistive layer halfwidth are shown as a function of $k$ for the static case. Similarly to the Harris sheet, the numerical matching quantity initially increases with $k$ before decreasing again. Interestingly though, the resistive layer halfwidth is decreasing as $k$ increases, but the product $\delta_\mathrm{N} |\Delta'_\mathrm{N}|$ never exceeds $\sim 0.20$, which would imply that the constant-$\psi$ approximation is valid across all $k$-values. However, as we concluded in Sec. \ref{sec:harris-delta}, the definition of $\Delta'_\mathrm{N}$ makes it unfit to compare to $\Delta'_\mathrm{A}$ in the nonconstant-$\psi$ regime, and ensures that $\delta_\mathrm{N} |\Delta'_\mathrm{N}| < 1$ is an insufficient criterion to classify a perturbation as a constant-$\psi$ mode. Additionally, though the growth rate variation with the wavenumber appears to follow a similar trend as the numerical matching quantity, it appears they are not directly proportional. This is further highlighted by Fig. \ref{fig:psi-rotation}(d), where the numerical matching quantity and resistive layer halfwidth are shown as a function of $v_\mathrm{c}$ for the stationary case. Here, the numerical matching quantity is observed to increase with $v_\mathrm{c}$ initially, whereas the growth rate starts to decline sooner as $v_\mathrm{c}$ increases. Here, the resistive layer halfwidth is observed to be more or less constant. Consequently, we henceforth abandon the study of the numerical matching quantity and its relation to the growth rate for this configuration, and solely focus on the modification of the growth rate by the flow profile.

\subsubsection{Growth rate scaling with resistivity}\label{sec:rot-eta}
\begin{figure*}
    \centering
    \includegraphics[width=\textwidth]{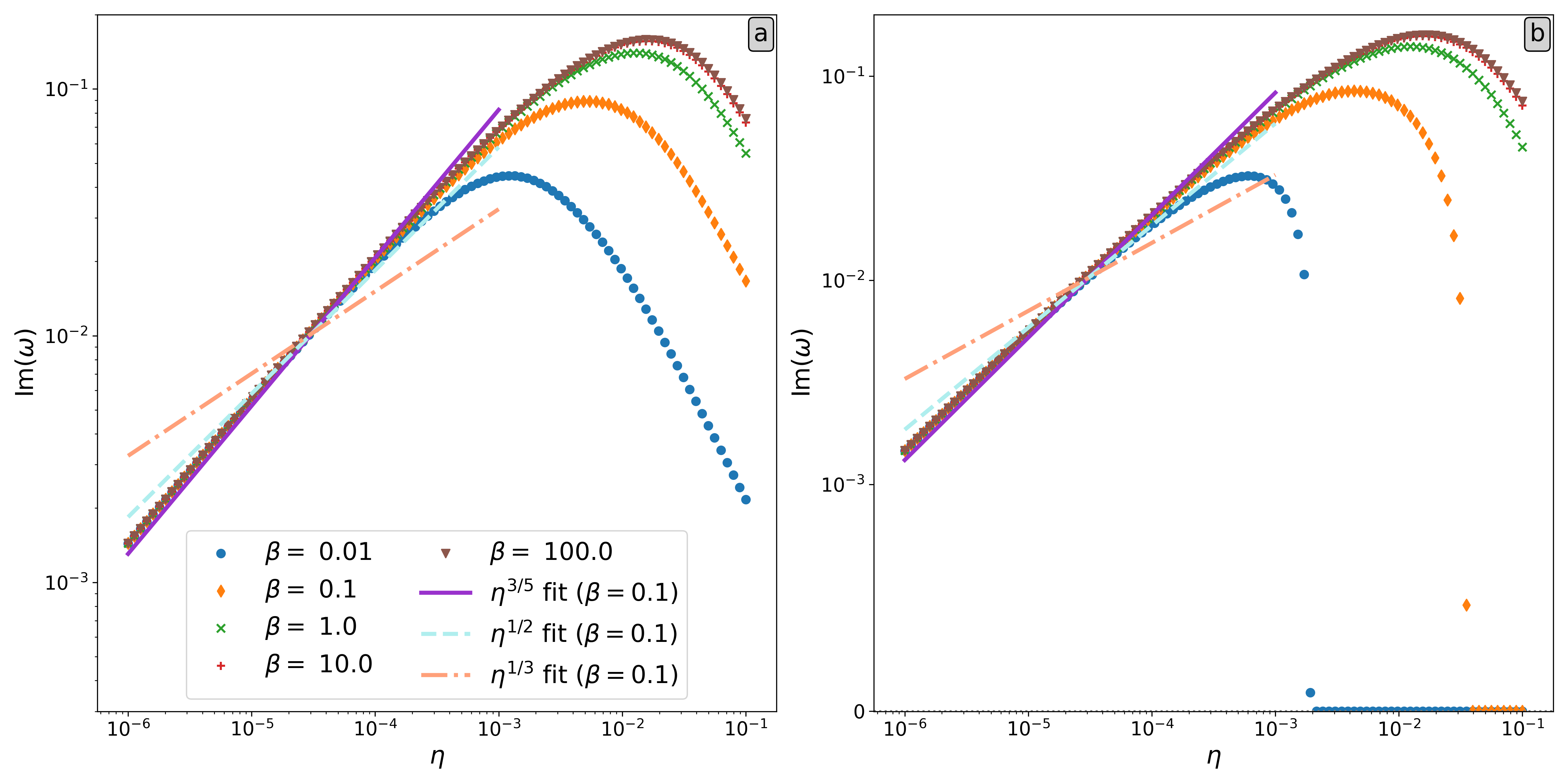}
    \caption{Resistive tearing growth rate of a force-free magnetic field, Eqs. (\ref{eq:equil-rotB}), for $\rho_\mathrm{c} = 1$ and $\bfk = 1.5\,\ey$, as a function of $\eta$ for various plasma-$\beta$ (a) without flow and (b) for $v_\mathrm{c} = 0.15$. Both panels include three power law fits ($\eta^{3/5}$, $\eta^{1/2}$, and $\eta^{1/3}$) for the case with $\beta = 0.1$.}
    \label{fig:rot-etavar}
\end{figure*}

Once again, we now turn to a comparison of the growth rate scaling with analytic predictions. The parameters are the same as in the previous section, except that $\beta$ takes on various values and $v_\mathrm{c} = 0.15$ when flow is included. As demonstrated, for this equilibrium we are dealing with a constant-$\psi$ mode and thus expect a scaling of $\text{Im}(\omega) \sim \eta^{3/5}$ in the absence of flow. For the case with flow, we expect an identical scaling since $R_0 \simeq 0.03 \ll 1$. \cite{Chen1990} This scaling is expected to hold for $a / \eta > 10^2$, since the boundary layer approach used in analytic works is no longer appropriate as the resistivity increases,\cite{Betar2022} and thus, substituting a quarter period for the transition halfwidth $a = \pi/2\alpha$, for $\eta <  3.3 \times 10^{-3}$. Hence, all comparisons were limited to $\eta < 10^{-3}$. Indeed, initially, the growth rate scales as $\eta^{3/5}$, as shown in Fig. \ref{fig:rot-etavar}(a) for $301$ grid points, independently of $\beta$. However, as $\eta$ approaches $10^{-3}$, the growth rate starts to deviate from this scaling, with smaller $\beta$-values deviating sooner, and eventually decreases again. Adding velocity to this variation in resistivity steepens the growth rate dropoff for lower $\beta$-values, as evidenced by Fig. \ref{fig:rot-etavar}(b), going as far as eliminating the instability entirely. Of course, care should again be taken in the interpretation of growth rates above $\eta \sim 10^{-2}$, since diffusion of the equilibrium was neglected.

\subsubsection{Density variation}
\begin{figure*}
    \centering
    \includegraphics[width=\textwidth]{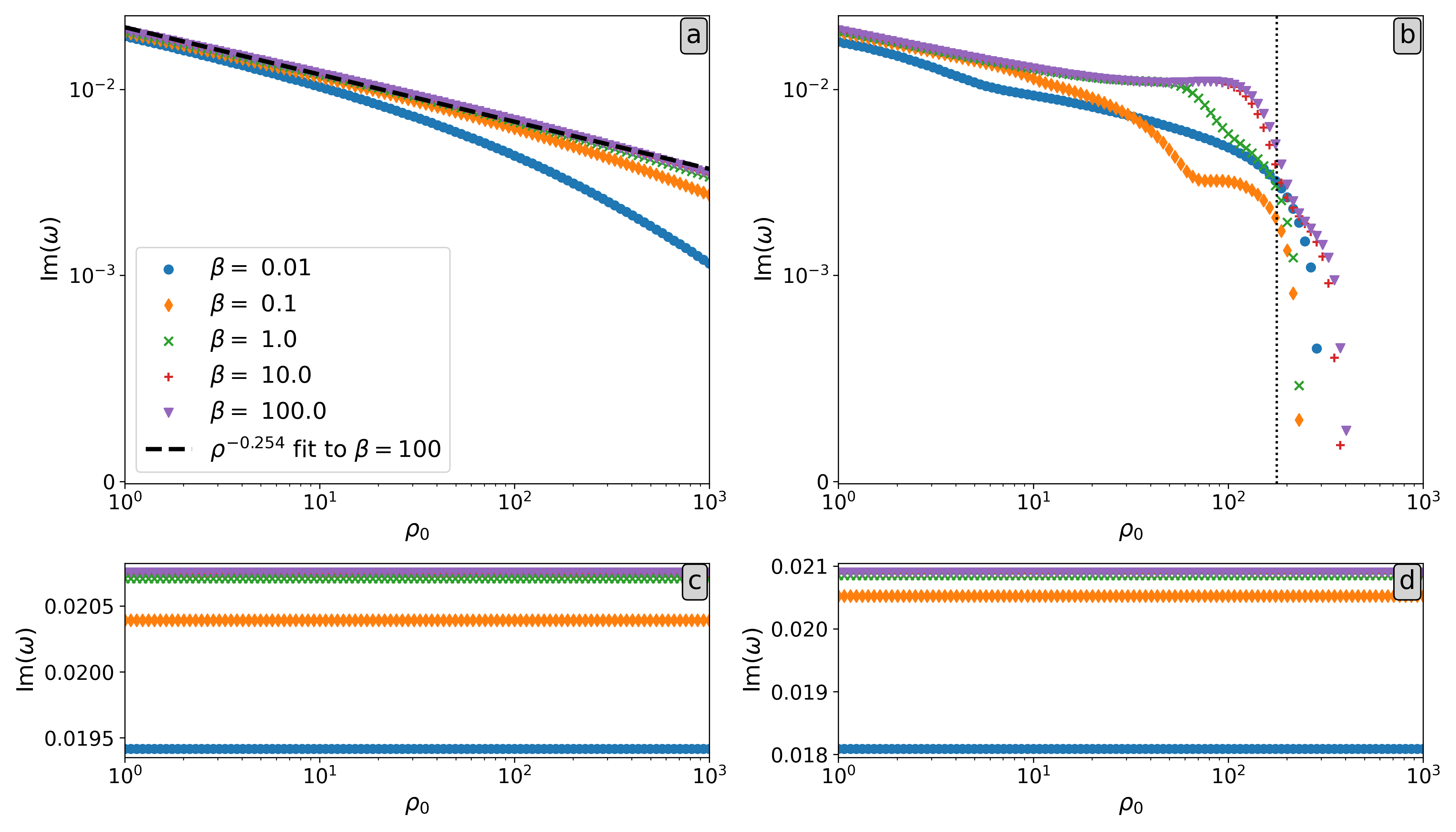}
    \caption{Resistive tearing growth rate of a force-free magnetic field, Eqs. (\ref{eq:equil-rotB}), for $\bfk = 1.5\,\ey$, as a function of $\rho_0$ for various plasma-$\beta$ (a) without flow and (b) for $v_\mathrm{c} = 0.15$. The dotted vertical line indicates where the maximal equilibrium speed equals the Alfv\'en speed. (c) and (d) show the growth rate for the same parameters as in (a) and (b), respectively, but with $\bfb_0$ scaled with $\sqrt{\rho_0}$ such that the Alfv\'en speed is constant.}
    \label{fig:rot-rhovar}
\end{figure*}

Adopting the same approach as for the Harris sheet configuration, we here show that the growth rate of the force-free magnetic field configuration also varies with the Alfv\'en speed. To do so, we first consider the flowless case with parameters $\bfk = 1.5\,\ey$ and $\alpha = 4.73884$, for various values of $\beta$ and varying the density $\rho_0$ ($301$ grid points). As can be seen in Fig. \ref{fig:rot-rhovar}(a), a similar growth rate scaling, $\rho_0^{-0.254}$, to the Harris sheet is recovered for high plasma-$\beta$. For this configuration the deviation from the analytic $\rho_0^{-1/5}$ scaling is even greater, though an explanation might be that this is due to the incompressible approximation breaking down at high $\beta$. However, as $\beta$ decreases, the growth rate deviates even more from a power law, especially for larger densities (smaller Alfv\'en speeds). This may be due to the inclusion of Joule heating, whose contribution to the energy equation is more significant for small $\beta$ ($T_0 \propto \beta$).

Now adding a velocity with $v_\mathrm{c} = 0.15$, behaviour similar to the Harris sheet is observed in Fig. \ref{fig:rot-rhovar}(b). Again, deviations from the flowless scaling are observed as the density approaches the critical value where the maximal flow in the domain equals the Alfv\'en speed (indicated with a dotted line). Above this threshold, the tearing mode is heavily damped and the growth rate goes to zero.

To show once again that this scaling is due to the change in Alfv\'en velocity, consider the above case but with $\bfb_0$ the profile in Eqs. (\ref{eq:equil-rotB}) multiplied with $\sqrt{\rho_0}$, such that the Alfv\'en speed is constant. The resulting growth rates are shown in Figs. \ref{fig:rot-rhovar}(c) and (d), for the static and stationary cases, respectively. In both cases, the growth rate is constant for all densities, with higher $\beta$ resulting in a higher growth rate. Hence, the growth rate does not depend on the specific density, but on the Alfv\'en speed. Since $\beta$ appears to affect the growth rate, its role is further investigated in conjunction with the velocity.

\subsubsection{Velocity variation}
\begin{figure*}
    \centering
    \includegraphics[width=\textwidth]{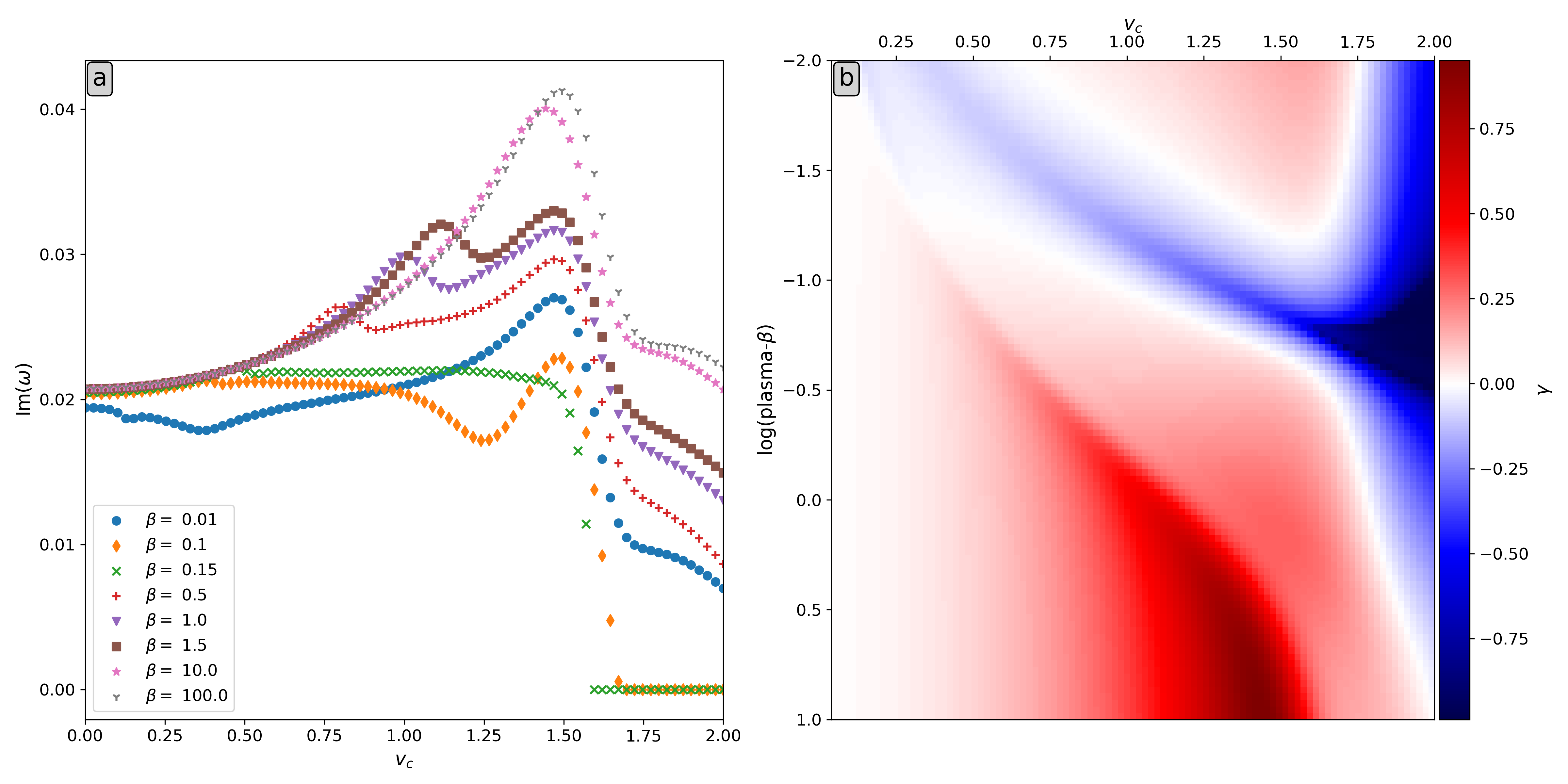}
    \caption{Resistive tearing mode growth rate of a force-free magnetic field, Eqs. (\ref{eq:equil-rotB}), for $\rho_\mathrm{c} = 1$. (a) Absolute growth rate as a function of $v_\mathrm{c}$ for various plasma-$\beta$ values. (b) Relative growth rate for varying $v_\mathrm{c}$ and $\beta$ with respect to the flowless growth rate.}
    \label{fig:rot-v02beta}
\end{figure*}

Despite the simple velocity profile, the $(v_\mathrm{c},\beta)$-parameter space reveals surprising complexity. After assuming a constant density $\rho_0 = 1$, the present parametric survey varied $v_\mathrm{c}$ and $\beta$ simultaneously for wave vector $\bfk = 1.5\,\ey$, where the $v_\mathrm{c}$ parameter was limited to the interval $[0, 2]$, such that the equilibrium velocity remains sub-Alfv\'enic ($|\bfv_0| \leq c_\mathrm{A}$) on the entire domain, and $\beta$-values from $10^{-2}$ to $10^2$ were studied. The results are shown in Fig. \ref{fig:rot-v02beta}, where all runs in panel (a) were performed at $301$ grid points, whereas $201$ grid points were used in panel (b).

As pointed out in Ref. \onlinecite{Hofmann1975}, the introduction of flow in a system that is unstable to the resistive tearing mode can either stabilise or further destabilise the plasma. This is also immediately clear from Fig. \ref{fig:rot-v02beta}(b), where blue indicates a stabilised system and red a strong increase in tearing growth rate. Whilst the plasma is mostly destabilised further by the presence of flow for large $\beta$, as clearly evidenced by Fig. \ref{fig:rot-v02beta}, the destabilising effect does not scale monotonically with the velocity coefficient. Rather, the maximal destabilisation appears at some intermediate value between small speeds and the Alfv\'en speed. For small to intermediate $\beta$ ($\lesssim 1$), on the other hand, both stabilising and destabilising influences are observed in significant fractions of the velocity space, with the strongest stabilising effect occurring at the Alfv\'en speed. Additionally, more than one stabilising-destabilising transition is observed in panel (b) along the speed axis for small $\beta$ ($\ll 1$).

This dependence on the plasma-$\beta$, which already appeared in a less pronounced way in Sec. \ref{sec:rot-eta}, is not surprising. The ion sound Larmor radius $\rho_\mathrm{s}$ is known to affect the growth rate,\cite{Betar2020,Betar2022} and $\beta$ enters in the radius through our definition of $T_0$. Nevertheless, any dependencies on $\rho_\mathrm{s}$ were derived using an isothermal closure, contrary to the equations used here. Since it is clear that the plasma-$\beta$ affects the role of flow significantly, a more in-depth analysis of the influence of $\beta$ on tearing modes in general offers interesting perspectives for future work.

\section{Conclusion}
In this work we studied the linear regime of the resistive tearing mode in a compressible plasma with Joule heating in two different configurations featuring shear flow: a Harris current sheet and a force-free magnetic field varying its direction periodically throughout a plasma slab.

First, we visualised the magnetic field lines and flow patterns of the linearly perturbed Harris sheet in 2D, both in the absence and presence of a background flow. In either case, the magnetic field lines were pinched together periodically along the sheet to reconnect and form magnetic islands, as observed in simulations. If this perturbation is allowed to evolve linearly, the formation of two smaller sub-islands is observed inside the island. Since this behaviour does not occur in non-linear simulations, the time where this formation is initiated in the linear evolution places an upper limit on the transition time from the linear to the non-linear regime.

For the velocity, the largest perturbation occurs at the magnetic nullplane for both cases with and without equilibrium flow, with plasma leaving the pinched regions and streaming towards the magnetic islands. The plasma further away from the nullplane has an almost negligible velocity compared to the plasma at the nullplane. However, for the flowless equilibrium the plasma was observed to move away from the magnetic islands here, whereas the plasma rotates inside the inner islands if a background flow is included.

Next, we introduced a numerical equivalent of the matching quantity $\Delta'$ and resistive layer halfwidth $\delta$. However, for both configurations the numerical quantities deviated from the analytic predictions. In the nonconstant-$\psi$ regime in particular, the numerical matching quantity fails to capture the behaviour of the analytic matching quantity, due to the lack of sharp transition in $\psi'/\psi$ near the nullplane for wider resistive layers. In addition, no clear scaling of the growth rate with the numerical matching quantity was observed. Therefore, we advocate against the use of the matching quantity as the sole quantifier of flow's influence on the tearing growth rate.

As a consequence of this deviation from analytic predictions in the matching quantity in the nonconstant-$\psi$ regime, the product $\delta_\mathrm{N} |\Delta'_\mathrm{N}|$ is not a good indicator of the validity of the constant-$\psi$ approximation. Nevertheless, the literature's scaling laws hold in the constant-$\psi$ regime, as far as we have observed. Indeed, the growth rates of both configurations were found to scale as $\eta^{3/5}$ in the absence of flow, as expected for a constant-$\psi$ mode. Subsequently, when flow was introduced, the transition to the constant-$\psi$ $\eta^{1/2}$ scaling was also observed for considerable flow speeds. However, in the case of the Harris sheet, a different scaling was found for a case with even larger flow speed. Of course, the growth rate scaling with resistivity deviated from the analytic scaling for stronger resistivities, as is to be expected, especially in the presence of flow. In both cases, flow was observed to introduce a cutoff resistivity above which the tearing mode is damped. In the case of the Harris sheet, this cutoff even lay inside the regime where the analytic scaling law is still expected to hold, though this was for the case that did not follow the analytic scaling law anyway.

Afterwards, we showed that the growth rate scales with the Alfv\'en speed, though the scaling was found to be closer to $c_\mathrm{A}^{1/2}$ than to the analytic $c_\mathrm{A}^{2/5}$. Additionally, a significant deviation from this scaling was observed for larger densities, i.e. smaller Alfv\'en speeds, in the force-free field case with small plasma-$\beta$. The importance of the plasma-$\beta$ was then further highlighted in this configuration, where an intricate interplay between the plasma-$\beta$ and flow speed was observed. For large plasma-$\beta$, the flow was observed to destabilise the plasma further, with the strongest destabilisation occurring at intermediate flow speeds. For small plasma-$\beta$ though, both stabilising and destabilising influences were observed, with the strongest stabilisation occurring at the Alfv\'en speed. Additionally, more than one stabilising-destabilising transition was observed along the flow speed axis for small plasma-$\beta$. Presumably, the importance of the plasma-$\beta$ in this study is due to the inclusion of Joule heating, which is absent in preceding work.

Finally, it was shown that the addition of flow to an equilibrium with multiple magnetic nullplanes and tearing modes can modify the tearing behaviour significantly. The tearing behaviour of all modes was altered, with all modes tearing all nullplanes, albeit to different degrees. This is in stark contrast to the static case, where only the dominant mode tore all nullplanes, and the secondary and tertiary modes only tore the central and non-central nullplanes, respectively.

Looking ahead, future work could focus more on the role of the plasma-$\beta$, especially at small values. Additionally, \Legolas{} could be used to incorporate viscosity,\cite{DeJonghe2022} or to investigate transitions in instability dominance, from resistive tearing to the Kelvin-Helmholtz instability, at near-Alfv\'enic speeds. Also the Hall field\cite{Shi2020} and electron inertia effects\cite{DeJonghe2022} are prime candidates for further investigation. Of course, a similar study to this one can be performed for the cylindrical tearing mode,\cite{Coppi1966} to compare the effects of axial and azimuthal flow. Finally, due to the growth rate depending on the specific flow profile, \Legolas{} could also be employed as a computationally inexpensive, diagnostic tool for concrete configurations, particularly for experiments and for comparison of linear theory to non-linear simulations.

\begin{acknowledgments}
This work was supported by funding from the European Research Council (ERC) under the European Unions Horizon 2020 research and innovation programme, Grant agreement No. 833251 PROMINENT ERC-ADG 2018. JDJ acknowledges further funding by the UK's Science and Technology Facilities Council (STFC) Consolidated Grant ST/W001195/1. RK is further supported by Internal Funds KU Leuven through the project C14/19/089 TRACESpace and an FWO project G0B4521N.

We thank the referees for their constructive and thought-provoking comments, which helped to improve the manuscript significantly.

The authors have no conflicts to disclose.
\end{acknowledgments}

\section*{Data Availability Statement}
The data that support the findings of this study are available from the corresponding author upon reasonable request.

All functionalities to reproduce the results presented here are available in \Legolas{} v$2.1.1$. The \Legolas{} code is freely available under the GNU General Public License. For more information, visit \url{https://legolas.science/}.

\appendix
\section{Accumulated grid}\label{app:grid}
Due to the heavily localised transitions in the equilibrium profiles of the Harris current sheet, an equidistant grid would require many more grid points than a centrally-accumulated grid to properly resolve the region of steepest change in the middle. Therefore, this study opted for a grid constructed using the algorithm below\cite{Quarteroni2009} for an interval $x\in[a,b]$ and function $f$.

\begin{algorithmic}
    \State Declare array $\mathrm{auxGrid}$
    \State Declare array $\mathrm{finalGrid}$
    \State $\mathrm{auxGrid}(1) \gets a$
    \State $i \gets 1$
    \While{$\mathrm{auxGrid}(i) < (a+b)/2$}
        \State $\mathrm{auxGrid}(i+1) \gets \mathrm{auxGrid}(i) + f(\mathrm{auxGrid}(i))$
        \State $i \gets i+1$
    \EndWhile
    \State $\mathrm{finalGrid}(1) \gets a$
    \State $\mathrm{finalGrid}(2i-1) \gets b$
    \State $\kappa \gets ((a+b)/2 - \mathrm{auxGrid}(i-1)) / (\mathrm{auxGrid}(i) - \mathrm{auxGrid}(i-1))$
    \For{$j$ from $1$ to $i-1$}
        \State $\mathrm{finalGrid}(j+1) \gets \mathrm{auxGrid}(j) + \kappa f(\mathrm{auxGrid}(j))$
        \State $\mathrm{finalGrid}(2i-j-1) \gets a + b - \mathrm{finalGrid}(j+1)$
    \EndFor
    \State $\mathrm{finalGrid}(i+1) \gets (a+b)/2$
\end{algorithmic}

This results in a symmetric grid of $2i-1$ grid points around the centre of the interval, $(a+b)/2$. A symmetric grid is desired because it reduces the likelihood that the spectrum's symmetry, in the case of an odd flow profile, is broken by numerical errors. In our specific case, we used the Gaussian function
\begin{equation}
    f(x) = p_1 - (p_1 - p_3)\,\exp\left( \frac{-(x-p_2)^2}{2p_4} \right)
\end{equation}
with $p_2 = 0$ to obtain a grid strongly accumulated around the interval's centre $0$.

\nocite{*}
\bibliography{bibliography.bib}

\begin{thebibliography}{64}%
\makeatletter
\providecommand \@ifxundefined [1]{%
 \@ifx{#1\undefined}
}%
\providecommand \@ifnum [1]{%
 \ifnum #1\expandafter \@firstoftwo
 \else \expandafter \@secondoftwo
 \fi
}%
\providecommand \@ifx [1]{%
 \ifx #1\expandafter \@firstoftwo
 \else \expandafter \@secondoftwo
 \fi
}%
\providecommand \natexlab [1]{#1}%
\providecommand \enquote  [1]{``#1''}%
\providecommand \bibnamefont  [1]{#1}%
\providecommand \bibfnamefont [1]{#1}%
\providecommand \citenamefont [1]{#1}%
\providecommand \href@noop [0]{\@secondoftwo}%
\providecommand \href [0]{\begingroup \@sanitize@url \@href}%
\providecommand \@href[1]{\@@startlink{#1}\@@href}%
\providecommand \@@href[1]{\endgroup#1\@@endlink}%
\providecommand \@sanitize@url [0]{\catcode `\\12\catcode `\$12\catcode `\&12\catcode `\#12\catcode `\^12\catcode `\_12\catcode `\%12\relax}%
\providecommand \@@startlink[1]{}%
\providecommand \@@endlink[0]{}%
\providecommand \url  [0]{\begingroup\@sanitize@url \@url }%
\providecommand \@url [1]{\endgroup\@href {#1}{\urlprefix }}%
\providecommand \urlprefix  [0]{URL }%
\providecommand \Eprint [0]{\href }%
\providecommand \doibase [0]{http://dx.doi.org/}%
\providecommand \selectlanguage [0]{\@gobble}%
\providecommand \bibinfo  [0]{\@secondoftwo}%
\providecommand \bibfield  [0]{\@secondoftwo}%
\providecommand \translation [1]{[#1]}%
\providecommand \BibitemOpen [0]{}%
\providecommand \bibitemStop [0]{}%
\providecommand \bibitemNoStop [0]{.\EOS\space}%
\providecommand \EOS [0]{\spacefactor3000\relax}%
\providecommand \BibitemShut  [1]{\csname bibitem#1\endcsname}%
\let\auto@bib@innerbib\@empty
\bibitem [{\citenamefont {{L\"orin\v{c}\'ik}}\ \emph {et~al.}(2021)\citenamefont {{L\"orin\v{c}\'ik}}, \citenamefont {{Dud\'ik}}, \citenamefont {Aulanier}, \citenamefont {Schmieder},\ and\ \citenamefont {Golub}}]{Lorincik2021}%
  \BibitemOpen
  \bibfield  {author} {\bibinfo {author} {\bibfnamefont {J.}~\bibnamefont {{L\"orin\v{c}\'ik}}}, \bibinfo {author} {\bibfnamefont {J.}~\bibnamefont {{Dud\'ik}}}, \bibinfo {author} {\bibfnamefont {G.}~\bibnamefont {Aulanier}}, \bibinfo {author} {\bibfnamefont {B.}~\bibnamefont {Schmieder}}, \ and\ \bibinfo {author} {\bibfnamefont {L.}~\bibnamefont {Golub}},\ }\bibfield  {title} {\enquote {\bibinfo {title} {Imaging evidence for solar wind outflows originating from a coronal mass ejection footpoint},}\ }\href {\doibase 10.3847/1538-4357/abc8f6} {\bibfield  {journal} {\bibinfo  {journal} {Astrophys. J.}\ }\textbf {\bibinfo {volume} {906}},\ \bibinfo {pages} {62} (\bibinfo {year} {2021})}\BibitemShut {NoStop}%
\bibitem [{\citenamefont {Phan}\ \emph {et~al.}(2022)\citenamefont {Phan}, \citenamefont {Verniero}, \citenamefont {Larson}, \citenamefont {Lavraud}, \citenamefont {Drake}, \citenamefont {{{\O}ieroset}}, \citenamefont {Eastwood}, \citenamefont {Bale}, \citenamefont {Livi}, \citenamefont {Halekas}, \citenamefont {Whittlesey}, \citenamefont {Rahmati}, \citenamefont {Stansby}, \citenamefont {Pulupa}, \citenamefont {MacDowall}, \citenamefont {Szabo}, \citenamefont {Koval}, \citenamefont {Desai}, \citenamefont {Fuselier}, \citenamefont {Velli}, \citenamefont {Hesse}, \citenamefont {Pyakurel}, \citenamefont {Maheshwari}, \citenamefont {Kasper}, \citenamefont {Stevens}, \citenamefont {Case},\ and\ \citenamefont {Raouafi}}]{Phan2022}%
  \BibitemOpen
  \bibfield  {author} {\bibinfo {author} {\bibfnamefont {T.~D.}\ \bibnamefont {Phan}}, \bibinfo {author} {\bibfnamefont {J.~L.}\ \bibnamefont {Verniero}}, \bibinfo {author} {\bibfnamefont {D.}~\bibnamefont {Larson}}, \bibinfo {author} {\bibfnamefont {B.}~\bibnamefont {Lavraud}}, \bibinfo {author} {\bibfnamefont {J.~F.}\ \bibnamefont {Drake}}, \bibinfo {author} {\bibfnamefont {M.}~\bibnamefont {{{\O}ieroset}}}, \bibinfo {author} {\bibfnamefont {J.~P.}\ \bibnamefont {Eastwood}}, \bibinfo {author} {\bibfnamefont {S.~D.}\ \bibnamefont {Bale}}, \bibinfo {author} {\bibfnamefont {R.}~\bibnamefont {Livi}}, \bibinfo {author} {\bibfnamefont {J.~S.}\ \bibnamefont {Halekas}}, \bibinfo {author} {\bibfnamefont {P.~L.}\ \bibnamefont {Whittlesey}}, \bibinfo {author} {\bibfnamefont {A.}~\bibnamefont {Rahmati}}, \bibinfo {author} {\bibfnamefont {D.}~\bibnamefont {Stansby}}, \bibinfo {author} {\bibfnamefont {M.}~\bibnamefont {Pulupa}}, \bibinfo {author} {\bibfnamefont {R.~J.}\ \bibnamefont {MacDowall}}, \bibinfo {author} {\bibfnamefont {P.~A.}\ \bibnamefont {Szabo}}, \bibinfo {author} {\bibfnamefont {A.}~\bibnamefont {Koval}}, \bibinfo {author} {\bibfnamefont {M.}~\bibnamefont {Desai}}, \bibinfo {author} {\bibfnamefont {S.~A.}\ \bibnamefont {Fuselier}}, \bibinfo {author} {\bibfnamefont {M.}~\bibnamefont {Velli}}, \bibinfo {author} {\bibfnamefont {M.}~\bibnamefont {Hesse}}, \bibinfo {author} {\bibfnamefont {P.~S.}\ \bibnamefont {Pyakurel}}, \bibinfo {author} {\bibfnamefont {K.}~\bibnamefont {Maheshwari}}, \bibinfo {author} {\bibfnamefont {J.~C.}\ \bibnamefont {Kasper}}, \bibinfo {author} {\bibfnamefont {J.~M.}\ \bibnamefont {Stevens}}, \bibinfo {author} {\bibfnamefont {A.~W.}\ \bibnamefont {Case}}, \ and\ \bibinfo {author} {\bibfnamefont {N.~E.}\ \bibnamefont {Raouafi}},\ }\bibfield  {title} {\enquote {\bibinfo {title} {{Parker Solar Probe} observations of solar wind energetic proton beams produced by magnetic reconnection in the near-sun heliospheric current sheet},}\ }\href {\doibase https://doi.org/10.1029/2021GL096986} {\bibfield  {journal} {\bibinfo  {journal} {Geophys. Res. Lett.}\ }\textbf {\bibinfo {volume} {49}},\ \bibinfo {pages} {e2021GL096986} (\bibinfo {year} {2022})}\BibitemShut {NoStop}%
\bibitem [{\citenamefont {Qi}\ \emph {et~al.}(2022)\citenamefont {Qi}, \citenamefont {Li}, \citenamefont {Russell}, \citenamefont {Ergun}, \citenamefont {Jia},\ and\ \citenamefont {Hubbert}}]{Qi2022}%
  \BibitemOpen
  \bibfield  {author} {\bibinfo {author} {\bibfnamefont {Y.}~\bibnamefont {Qi}}, \bibinfo {author} {\bibfnamefont {T.~C.}\ \bibnamefont {Li}}, \bibinfo {author} {\bibfnamefont {C.~T.}\ \bibnamefont {Russell}}, \bibinfo {author} {\bibfnamefont {R.~E.}\ \bibnamefont {Ergun}}, \bibinfo {author} {\bibfnamefont {Y.-D.}\ \bibnamefont {Jia}}, \ and\ \bibinfo {author} {\bibfnamefont {M.}~\bibnamefont {Hubbert}},\ }\bibfield  {title} {\enquote {\bibinfo {title} {Magnetic flux transport identification of active reconnection: {MMS} observations in {Earth}’s magnetosphere},}\ }\href {\doibase 10.3847/2041-8213/ac5181} {\bibfield  {journal} {\bibinfo  {journal} {Astrophys. J. Lett.}\ }\textbf {\bibinfo {volume} {926}},\ \bibinfo {pages} {L34} (\bibinfo {year} {2022})}\BibitemShut {NoStop}%
\bibitem [{\citenamefont {{Sweet}}(1958)}]{Sweet1958}%
  \BibitemOpen
  \bibfield  {author} {\bibinfo {author} {\bibfnamefont {P.~A.}\ \bibnamefont {{Sweet}}},\ }\bibfield  {title} {\enquote {\bibinfo {title} {{The Neutral Point Theory of Solar Flares}},}\ }in\ \href@noop {} {\emph {\bibinfo {booktitle} {Electromagnetic Phenomena in Cosmical Physics}}},\ Vol.~\bibinfo {volume} {6},\ \bibinfo {editor} {edited by\ \bibinfo {editor} {\bibfnamefont {B.}~\bibnamefont {{Lehnert}}}}\ (\bibinfo {year} {1958})\ p.\ \bibinfo {pages} {123}\BibitemShut {NoStop}%
\bibitem [{\citenamefont {{Parker}}(1957)}]{Parker1957}%
  \BibitemOpen
  \bibfield  {author} {\bibinfo {author} {\bibfnamefont {E.~N.}\ \bibnamefont {{Parker}}},\ }\bibfield  {title} {\enquote {\bibinfo {title} {{Sweet's Mechanism for Merging Magnetic Fields in Conducting Fluids}},}\ }\href {\doibase 10.1029/JZ062i004p00509} {\bibfield  {journal} {\bibinfo  {journal} {J. Geophys. Res.}\ }\textbf {\bibinfo {volume} {62}},\ \bibinfo {pages} {509--520} (\bibinfo {year} {1957})}\BibitemShut {NoStop}%
\bibitem [{\citenamefont {{Petschek}}(1964)}]{Petschek1964}%
  \BibitemOpen
  \bibfield  {author} {\bibinfo {author} {\bibfnamefont {H.~E.}\ \bibnamefont {{Petschek}}},\ }\bibfield  {title} {\enquote {\bibinfo {title} {{Magnetic Field Annihilation}},}\ }in\ \href@noop {} {\emph {\bibinfo {booktitle} {NASA Special Publication}}},\ Vol.~\bibinfo {volume} {50}\ (\bibinfo {year} {1964})\ p.\ \bibinfo {pages} {425}\BibitemShut {NoStop}%
\bibitem [{\citenamefont {Furth}, \citenamefont {Killeen},\ and\ \citenamefont {Rosenbluth}(1963)}]{Furth1963}%
  \BibitemOpen
  \bibfield  {author} {\bibinfo {author} {\bibfnamefont {H.~P.}\ \bibnamefont {Furth}}, \bibinfo {author} {\bibfnamefont {J.}~\bibnamefont {Killeen}}, \ and\ \bibinfo {author} {\bibfnamefont {M.~N.}\ \bibnamefont {Rosenbluth}},\ }\bibfield  {title} {\enquote {\bibinfo {title} {Finite‐resistivity instabilities of a sheet pinch},}\ }\href {\doibase 10.1063/1.1706761} {\bibfield  {journal} {\bibinfo  {journal} {Phys. Fluids}\ }\textbf {\bibinfo {volume} {6}},\ \bibinfo {pages} {459--484} (\bibinfo {year} {1963})}\BibitemShut {NoStop}%
\bibitem [{\citenamefont {Birn}\ \emph {et~al.}(2001)\citenamefont {Birn}, \citenamefont {Drake}, \citenamefont {Shay}, \citenamefont {Rogers}, \citenamefont {Denton}, \citenamefont {Hesse}, \citenamefont {Kuznetsova}, \citenamefont {Ma}, \citenamefont {Bhattacharjee}, \citenamefont {Otto},\ and\ \citenamefont {Pritchett}}]{Birn2001}%
  \BibitemOpen
  \bibfield  {author} {\bibinfo {author} {\bibfnamefont {J.}~\bibnamefont {Birn}}, \bibinfo {author} {\bibfnamefont {J.~F.}\ \bibnamefont {Drake}}, \bibinfo {author} {\bibfnamefont {M.~A.}\ \bibnamefont {Shay}}, \bibinfo {author} {\bibfnamefont {B.~N.}\ \bibnamefont {Rogers}}, \bibinfo {author} {\bibfnamefont {R.~E.}\ \bibnamefont {Denton}}, \bibinfo {author} {\bibfnamefont {M.}~\bibnamefont {Hesse}}, \bibinfo {author} {\bibfnamefont {M.}~\bibnamefont {Kuznetsova}}, \bibinfo {author} {\bibfnamefont {Z.~W.}\ \bibnamefont {Ma}}, \bibinfo {author} {\bibfnamefont {A.}~\bibnamefont {Bhattacharjee}}, \bibinfo {author} {\bibfnamefont {A.}~\bibnamefont {Otto}}, \ and\ \bibinfo {author} {\bibfnamefont {P.~L.}\ \bibnamefont {Pritchett}},\ }\bibfield  {title} {\enquote {\bibinfo {title} {Geospace environmental modeling ({GEM}) magnetic reconnection challenge},}\ }\href {\doibase https://doi.org/10.1029/1999JA900449} {\bibfield  {journal} {\bibinfo  {journal} {J. Geophys. Res. Space Phys.}\ }\textbf {\bibinfo {volume} {106}},\ \bibinfo {pages} {3715--3719} (\bibinfo {year} {2001})}\BibitemShut {NoStop}%
\bibitem [{\citenamefont {{Terasawa}}(1983)}]{Terasawa1983}%
  \BibitemOpen
  \bibfield  {author} {\bibinfo {author} {\bibfnamefont {T.}~\bibnamefont {{Terasawa}}},\ }\bibfield  {title} {\enquote {\bibinfo {title} {{Hall current effect on tearing mode instability}},}\ }\href {\doibase 10.1029/GL010i006p00475} {\bibfield  {journal} {\bibinfo  {journal} {Geophys. Res. Lett.}\ }\textbf {\bibinfo {volume} {10}},\ \bibinfo {pages} {475--478} (\bibinfo {year} {1983})}\BibitemShut {NoStop}%
\bibitem [{\citenamefont {{Fruchtman}}\ and\ \citenamefont {{Strauss}}(1993)}]{Fruchtman1993}%
  \BibitemOpen
  \bibfield  {author} {\bibinfo {author} {\bibfnamefont {A.}~\bibnamefont {{Fruchtman}}}\ and\ \bibinfo {author} {\bibfnamefont {H.~R.}\ \bibnamefont {{Strauss}}},\ }\bibfield  {title} {\enquote {\bibinfo {title} {{Modification of short scale-length tearing modes by the Hall field}},}\ }\href {\doibase 10.1063/1.860880} {\bibfield  {journal} {\bibinfo  {journal} {Phys. Fluids B}\ }\textbf {\bibinfo {volume} {5}},\ \bibinfo {pages} {1408--1412} (\bibinfo {year} {1993})}\BibitemShut {NoStop}%
\bibitem [{\citenamefont {{Huba}}\ and\ \citenamefont {{Rudakov}}(2004)}]{Huba2004}%
  \BibitemOpen
  \bibfield  {author} {\bibinfo {author} {\bibfnamefont {J.~D.}\ \bibnamefont {{Huba}}}\ and\ \bibinfo {author} {\bibfnamefont {L.~I.}\ \bibnamefont {{Rudakov}}},\ }\bibfield  {title} {\enquote {\bibinfo {title} {{Hall Magnetic Reconnection Rate}},}\ }\href {\doibase 10.1103/PhysRevLett.93.175003} {\bibfield  {journal} {\bibinfo  {journal} {Phys. Rev. Lett.}\ }\textbf {\bibinfo {volume} {93}},\ \bibinfo {eid} {175003} (\bibinfo {year} {2004})}\BibitemShut {NoStop}%
\bibitem [{\citenamefont {{Pucci}}, \citenamefont {{Velli}},\ and\ \citenamefont {{Tenerani}}(2017)}]{Pucci2017}%
  \BibitemOpen
  \bibfield  {author} {\bibinfo {author} {\bibfnamefont {F.}~\bibnamefont {{Pucci}}}, \bibinfo {author} {\bibfnamefont {M.}~\bibnamefont {{Velli}}}, \ and\ \bibinfo {author} {\bibfnamefont {A.}~\bibnamefont {{Tenerani}}},\ }\bibfield  {title} {\enquote {\bibinfo {title} {{Fast Magnetic Reconnection: {\textquotedblleft}Ideal{\textquotedblright} Tearing and the Hall Effect}},}\ }\href {\doibase 10.3847/1538-4357/aa7b82} {\bibfield  {journal} {\bibinfo  {journal} {Astrophys. J.}\ }\textbf {\bibinfo {volume} {845}},\ \bibinfo {eid} {25} (\bibinfo {year} {2017})},\ \Eprint {http://arxiv.org/abs/1704.08793} {arXiv:1704.08793 [astro-ph.SR]} \BibitemShut {NoStop}%
\bibitem [{\citenamefont {{Papini}}, \citenamefont {{Landi}},\ and\ \citenamefont {{Del Zanna}}(2019)}]{Papini2019}%
  \BibitemOpen
  \bibfield  {author} {\bibinfo {author} {\bibfnamefont {E.}~\bibnamefont {{Papini}}}, \bibinfo {author} {\bibfnamefont {S.}~\bibnamefont {{Landi}}}, \ and\ \bibinfo {author} {\bibfnamefont {L.}~\bibnamefont {{Del Zanna}}},\ }\bibfield  {title} {\enquote {\bibinfo {title} {{Fast Magnetic Reconnection: Secondary Tearing Instability and Role of the Hall Term}},}\ }\href {\doibase 10.3847/1538-4357/ab4352} {\bibfield  {journal} {\bibinfo  {journal} {Astrophys. J.}\ }\textbf {\bibinfo {volume} {885}},\ \bibinfo {eid} {56} (\bibinfo {year} {2019})},\ \Eprint {http://arxiv.org/abs/1906.06779} {arXiv:1906.06779 [physics.plasm-ph]} \BibitemShut {NoStop}%
\bibitem [{\citenamefont {Shi}\ \emph {et~al.}(2020)\citenamefont {Shi}, \citenamefont {Velli}, \citenamefont {Pucci}, \citenamefont {Tenerani},\ and\ \citenamefont {Innocenti}}]{Shi2020}%
  \BibitemOpen
  \bibfield  {author} {\bibinfo {author} {\bibfnamefont {C.}~\bibnamefont {Shi}}, \bibinfo {author} {\bibfnamefont {M.}~\bibnamefont {Velli}}, \bibinfo {author} {\bibfnamefont {F.}~\bibnamefont {Pucci}}, \bibinfo {author} {\bibfnamefont {A.}~\bibnamefont {Tenerani}}, \ and\ \bibinfo {author} {\bibfnamefont {M.~E.}\ \bibnamefont {Innocenti}},\ }\bibfield  {title} {\enquote {\bibinfo {title} {Oblique tearing mode instability: Guide field and {Hall} effect},}\ }\href {\doibase 10.3847/1538-4357/abb6fa} {\bibfield  {journal} {\bibinfo  {journal} {Astrophys. J.}\ }\textbf {\bibinfo {volume} {902}},\ \bibinfo {pages} {142} (\bibinfo {year} {2020})}\BibitemShut {NoStop}%
\bibitem [{\citenamefont {{De Jonghe}}, \citenamefont {Claes},\ and\ \citenamefont {Keppens}(2022)}]{DeJonghe2022}%
  \BibitemOpen
  \bibfield  {author} {\bibinfo {author} {\bibfnamefont {J.}~\bibnamefont {{De Jonghe}}}, \bibinfo {author} {\bibfnamefont {N.}~\bibnamefont {Claes}}, \ and\ \bibinfo {author} {\bibfnamefont {R.}~\bibnamefont {Keppens}},\ }\bibfield  {title} {\enquote {\bibinfo {title} {{Legolas : magnetohydrodynamic spectroscopy with viscosity and Hall current}},}\ }\href {\doibase 10.1017/S0022377822000617} {\bibfield  {journal} {\bibinfo  {journal} {J. Plasma Phys.}\ }\textbf {\bibinfo {volume} {88}},\ \bibinfo {pages} {905880321} (\bibinfo {year} {2022})}\BibitemShut {NoStop}%
\bibitem [{\citenamefont {{Wang}}\ and\ \citenamefont {{Bhattacharjee}}(1993)}]{Wang1993}%
  \BibitemOpen
  \bibfield  {author} {\bibinfo {author} {\bibfnamefont {X.}~\bibnamefont {{Wang}}}\ and\ \bibinfo {author} {\bibfnamefont {A.}~\bibnamefont {{Bhattacharjee}}},\ }\bibfield  {title} {\enquote {\bibinfo {title} {{Nonlinear dynamics of the m=1 instability and fast sawtooth collapse in high-temperature plasmas}},}\ }\href {\doibase 10.1103/PhysRevLett.70.1627} {\bibfield  {journal} {\bibinfo  {journal} {Phys. Rev. Lett.}\ }\textbf {\bibinfo {volume} {70}},\ \bibinfo {pages} {1627--1630} (\bibinfo {year} {1993})}\BibitemShut {NoStop}%
\bibitem [{\citenamefont {{Kleva}}, \citenamefont {{Drake}},\ and\ \citenamefont {{Waelbroeck}}(1995)}]{Kleva1995}%
  \BibitemOpen
  \bibfield  {author} {\bibinfo {author} {\bibfnamefont {R.~G.}\ \bibnamefont {{Kleva}}}, \bibinfo {author} {\bibfnamefont {J.~F.}\ \bibnamefont {{Drake}}}, \ and\ \bibinfo {author} {\bibfnamefont {F.~L.}\ \bibnamefont {{Waelbroeck}}},\ }\bibfield  {title} {\enquote {\bibinfo {title} {{Fast reconnection in high temperature plasmas}},}\ }\href {\doibase 10.1063/1.871095} {\bibfield  {journal} {\bibinfo  {journal} {Phys. Plasmas}\ }\textbf {\bibinfo {volume} {2}},\ \bibinfo {pages} {23--34} (\bibinfo {year} {1995})}\BibitemShut {NoStop}%
\bibitem [{\citenamefont {{Wang}}, \citenamefont {{Bhattacharjee}},\ and\ \citenamefont {{Ma}}(2000)}]{Wang2000}%
  \BibitemOpen
  \bibfield  {author} {\bibinfo {author} {\bibfnamefont {X.}~\bibnamefont {{Wang}}}, \bibinfo {author} {\bibfnamefont {A.}~\bibnamefont {{Bhattacharjee}}}, \ and\ \bibinfo {author} {\bibfnamefont {Z.~W.}\ \bibnamefont {{Ma}}},\ }\bibfield  {title} {\enquote {\bibinfo {title} {{Collisionless reconnection: Effects of Hall current and electron pressure gradient}},}\ }\href {\doibase 10.1029/1999JA000357} {\bibfield  {journal} {\bibinfo  {journal} {J. Geophys. Res.}\ }\textbf {\bibinfo {volume} {105}},\ \bibinfo {pages} {27633--27648} (\bibinfo {year} {2000})}\BibitemShut {NoStop}%
\bibitem [{\citenamefont {{Cai}}\ and\ \citenamefont {{Lee}}(1997)}]{Cai1997}%
  \BibitemOpen
  \bibfield  {author} {\bibinfo {author} {\bibfnamefont {H.~J.}\ \bibnamefont {{Cai}}}\ and\ \bibinfo {author} {\bibfnamefont {L.~C.}\ \bibnamefont {{Lee}}},\ }\bibfield  {title} {\enquote {\bibinfo {title} {{The generalized Ohm's law in collisionless magnetic reconnection}},}\ }\href {\doibase 10.1063/1.872178} {\bibfield  {journal} {\bibinfo  {journal} {Phys. Plasmas}\ }\textbf {\bibinfo {volume} {4}},\ \bibinfo {pages} {509--520} (\bibinfo {year} {1997})}\BibitemShut {NoStop}%
\bibitem [{\citenamefont {{Yin}}\ \emph {et~al.}(2001)\citenamefont {{Yin}}, \citenamefont {{Winske}}, \citenamefont {{Gary}},\ and\ \citenamefont {{Birn}}}]{Yin2001}%
  \BibitemOpen
  \bibfield  {author} {\bibinfo {author} {\bibfnamefont {L.}~\bibnamefont {{Yin}}}, \bibinfo {author} {\bibfnamefont {D.}~\bibnamefont {{Winske}}}, \bibinfo {author} {\bibfnamefont {S.~P.}\ \bibnamefont {{Gary}}}, \ and\ \bibinfo {author} {\bibfnamefont {J.}~\bibnamefont {{Birn}}},\ }\bibfield  {title} {\enquote {\bibinfo {title} {{Hybrid and Hall-MHD simulations of collisionless reconnection: Dynamics of the electron pressure tensor}},}\ }\href {\doibase 10.1029/2000JA000398} {\bibfield  {journal} {\bibinfo  {journal} {J. Geophys. Res.}\ }\textbf {\bibinfo {volume} {106}},\ \bibinfo {pages} {10761--10776} (\bibinfo {year} {2001})}\BibitemShut {NoStop}%
\bibitem [{\citenamefont {{Mandt}}, \citenamefont {{Denton}},\ and\ \citenamefont {{Drake}}(1994)}]{Mandt1994}%
  \BibitemOpen
  \bibfield  {author} {\bibinfo {author} {\bibfnamefont {M.~E.}\ \bibnamefont {{Mandt}}}, \bibinfo {author} {\bibfnamefont {R.~E.}\ \bibnamefont {{Denton}}}, \ and\ \bibinfo {author} {\bibfnamefont {J.~F.}\ \bibnamefont {{Drake}}},\ }\bibfield  {title} {\enquote {\bibinfo {title} {{Transition to whistler mediated magnetic reconnection}},}\ }\href {\doibase 10.1029/93GL03382} {\bibfield  {journal} {\bibinfo  {journal} {Geophys. Res. Lett.}\ }\textbf {\bibinfo {volume} {21}},\ \bibinfo {pages} {73--76} (\bibinfo {year} {1994})}\BibitemShut {NoStop}%
\bibitem [{\citenamefont {{Shay}}\ \emph {et~al.}(2001)\citenamefont {{Shay}}, \citenamefont {{Drake}}, \citenamefont {{Rogers}},\ and\ \citenamefont {{Denton}}}]{Shay2001}%
  \BibitemOpen
  \bibfield  {author} {\bibinfo {author} {\bibfnamefont {M.~A.}\ \bibnamefont {{Shay}}}, \bibinfo {author} {\bibfnamefont {J.~F.}\ \bibnamefont {{Drake}}}, \bibinfo {author} {\bibfnamefont {B.~N.}\ \bibnamefont {{Rogers}}}, \ and\ \bibinfo {author} {\bibfnamefont {R.~E.}\ \bibnamefont {{Denton}}},\ }\bibfield  {title} {\enquote {\bibinfo {title} {{Alfv{\'e}nic collisionless magnetic reconnection and the Hall term}},}\ }\href {\doibase 10.1029/1999JA001007} {\bibfield  {journal} {\bibinfo  {journal} {J. Geophys. Res.}\ }\textbf {\bibinfo {volume} {106}},\ \bibinfo {pages} {3759--3772} (\bibinfo {year} {2001})}\BibitemShut {NoStop}%
\bibitem [{\citenamefont {{Liu}}\ \emph {et~al.}(2022)\citenamefont {{Liu}}, \citenamefont {{Cassak}}, \citenamefont {{Li}}, \citenamefont {{Hesse}}, \citenamefont {{Lin}},\ and\ \citenamefont {{Genestreti}}}]{Liu2022}%
  \BibitemOpen
  \bibfield  {author} {\bibinfo {author} {\bibfnamefont {Y.-H.}\ \bibnamefont {{Liu}}}, \bibinfo {author} {\bibfnamefont {P.}~\bibnamefont {{Cassak}}}, \bibinfo {author} {\bibfnamefont {X.}~\bibnamefont {{Li}}}, \bibinfo {author} {\bibfnamefont {M.}~\bibnamefont {{Hesse}}}, \bibinfo {author} {\bibfnamefont {S.-C.}\ \bibnamefont {{Lin}}}, \ and\ \bibinfo {author} {\bibfnamefont {K.}~\bibnamefont {{Genestreti}}},\ }\bibfield  {title} {\enquote {\bibinfo {title} {{First-principles theory of the rate of magnetic reconnection in magnetospheric and solar plasmas}},}\ }\href {\doibase 10.1038/s42005-022-00854-x} {\bibfield  {journal} {\bibinfo  {journal} {Commun. Phys.}\ }\textbf {\bibinfo {volume} {5}},\ \bibinfo {eid} {97} (\bibinfo {year} {2022})},\ \Eprint {http://arxiv.org/abs/2203.14268} {arXiv:2203.14268 [physics.plasm-ph]} \BibitemShut {NoStop}%
\bibitem [{\citenamefont {{Bhattacharjee}}(2004)}]{Bhattacharjee2004}%
  \BibitemOpen
  \bibfield  {author} {\bibinfo {author} {\bibfnamefont {A.}~\bibnamefont {{Bhattacharjee}}},\ }\bibfield  {title} {\enquote {\bibinfo {title} {{Impulsive Magnetic Reconnection in the Earth's Magnetotail and the Solar Corona}},}\ }\href {\doibase 10.1146/annurev.astro.42.053102.134039} {\bibfield  {journal} {\bibinfo  {journal} {Annu. Rev. Astron. Astrophys.}\ }\textbf {\bibinfo {volume} {42}},\ \bibinfo {pages} {365--384} (\bibinfo {year} {2004})}\BibitemShut {NoStop}%
\bibitem [{\citenamefont {{Daughton}}\ \emph {et~al.}(2009)\citenamefont {{Daughton}}, \citenamefont {{Roytershteyn}}, \citenamefont {{Albright}}, \citenamefont {{Karimabadi}}, \citenamefont {{Yin}},\ and\ \citenamefont {{Bowers}}}]{Daughton2009}%
  \BibitemOpen
  \bibfield  {author} {\bibinfo {author} {\bibfnamefont {W.}~\bibnamefont {{Daughton}}}, \bibinfo {author} {\bibfnamefont {V.}~\bibnamefont {{Roytershteyn}}}, \bibinfo {author} {\bibfnamefont {B.~J.}\ \bibnamefont {{Albright}}}, \bibinfo {author} {\bibfnamefont {H.}~\bibnamefont {{Karimabadi}}}, \bibinfo {author} {\bibfnamefont {L.}~\bibnamefont {{Yin}}}, \ and\ \bibinfo {author} {\bibfnamefont {K.~J.}\ \bibnamefont {{Bowers}}},\ }\bibfield  {title} {\enquote {\bibinfo {title} {{Transition from collisional to kinetic regimes in large-scale reconnection layers}},}\ }\href {\doibase 10.1103/PhysRevLett.103.065004} {\bibfield  {journal} {\bibinfo  {journal} {Phys. Rev. Lett.}\ }\textbf {\bibinfo {volume} {103}},\ \bibinfo {eid} {065004} (\bibinfo {year} {2009})}\BibitemShut {NoStop}%
\bibitem [{\citenamefont {{Coppi}}(1964)}]{Coppi1964}%
  \BibitemOpen
  \bibfield  {author} {\bibinfo {author} {\bibfnamefont {B.}~\bibnamefont {{Coppi}}},\ }\bibfield  {title} {\enquote {\bibinfo {title} {{{\textquotedblleft}Inertial{\textquotedblright} instabilities in plasmas}},}\ }\href {\doibase 10.1016/0031-9163(64)90419-6} {\bibfield  {journal} {\bibinfo  {journal} {Phys. Lett.}\ }\textbf {\bibinfo {volume} {11}},\ \bibinfo {pages} {226--228} (\bibinfo {year} {1964})}\BibitemShut {NoStop}%
\bibitem [{\citenamefont {Shibata}\ and\ \citenamefont {Tanuma}(2001)}]{Shibata2001}%
  \BibitemOpen
  \bibfield  {author} {\bibinfo {author} {\bibfnamefont {K.}~\bibnamefont {Shibata}}\ and\ \bibinfo {author} {\bibfnamefont {S.}~\bibnamefont {Tanuma}},\ }\bibfield  {title} {\enquote {\bibinfo {title} {Plasmoid-induced-reconnection and fractal reconnection},}\ }\href {\doibase 10.1186/BF03353258} {\bibfield  {journal} {\bibinfo  {journal} {Earth Planets Space}\ }\textbf {\bibinfo {volume} {53}},\ \bibinfo {pages} {473--482} (\bibinfo {year} {2001})}\BibitemShut {NoStop}%
\bibitem [{\citenamefont {Li}\ and\ \citenamefont {Ma}(2010)}]{Li2010}%
  \BibitemOpen
  \bibfield  {author} {\bibinfo {author} {\bibfnamefont {J.~H.}\ \bibnamefont {Li}}\ and\ \bibinfo {author} {\bibfnamefont {Z.~W.}\ \bibnamefont {Ma}},\ }\bibfield  {title} {\enquote {\bibinfo {title} {{Nonlinear evolution of resistive tearing mode with sub-{Alfv\'enic} shear flow}},}\ }\href {\doibase 10.1029/2010JA015315} {\bibfield  {journal} {\bibinfo  {journal} {J. Geophys. Res. Space Phys.}\ }\textbf {\bibinfo {volume} {115}},\ \bibinfo {pages} {6--11} (\bibinfo {year} {2010})}\BibitemShut {NoStop}%
\bibitem [{\citenamefont {Li}\ and\ \citenamefont {Ma}(2012)}]{Li2012}%
  \BibitemOpen
  \bibfield  {author} {\bibinfo {author} {\bibfnamefont {J.~H.}\ \bibnamefont {Li}}\ and\ \bibinfo {author} {\bibfnamefont {Z.~W.}\ \bibnamefont {Ma}},\ }\bibfield  {title} {\enquote {\bibinfo {title} {Roles of super-{Alfv\'enic} shear flows on {Kelvin–Helmholtz} and tearing instability in compressible plasma},}\ }\href {\doibase 10.1088/0031-8949/86/04/045503} {\bibfield  {journal} {\bibinfo  {journal} {Phys. Scr.}\ }\textbf {\bibinfo {volume} {86}},\ \bibinfo {pages} {045503} (\bibinfo {year} {2012})}\BibitemShut {NoStop}%
\bibitem [{\citenamefont {Hofmann}(1975)}]{Hofmann1975}%
  \BibitemOpen
  \bibfield  {author} {\bibinfo {author} {\bibfnamefont {I.}~\bibnamefont {Hofmann}},\ }\bibfield  {title} {\enquote {\bibinfo {title} {Resistive tearing modes in a sheet pinch with shear flow},}\ }\href {\doibase 10.1088/0032-1028/17/2/005} {\bibfield  {journal} {\bibinfo  {journal} {Plasma Phys.}\ }\textbf {\bibinfo {volume} {17}},\ \bibinfo {pages} {143--157} (\bibinfo {year} {1975})}\BibitemShut {NoStop}%
\bibitem [{\citenamefont {Pollard}\ and\ \citenamefont {Taylor}(1979)}]{Pollard1979}%
  \BibitemOpen
  \bibfield  {author} {\bibinfo {author} {\bibfnamefont {R.~K.}\ \bibnamefont {Pollard}}\ and\ \bibinfo {author} {\bibfnamefont {J.~B.}\ \bibnamefont {Taylor}},\ }\bibfield  {title} {\enquote {\bibinfo {title} {Influence of equilibrium flows on tearing modes},}\ }\href {\doibase 10.1063/1.862451} {\bibfield  {journal} {\bibinfo  {journal} {Phys. Fluids}\ }\textbf {\bibinfo {volume} {22}},\ \bibinfo {pages} {126--131} (\bibinfo {year} {1979})}\BibitemShut {NoStop}%
\bibitem [{\citenamefont {Paris}\ and\ \citenamefont {Sy}(1983)}]{Paris1983}%
  \BibitemOpen
  \bibfield  {author} {\bibinfo {author} {\bibfnamefont {R.~B.}\ \bibnamefont {Paris}}\ and\ \bibinfo {author} {\bibfnamefont {W.~N.}\ \bibnamefont {Sy}},\ }\bibfield  {title} {\enquote {\bibinfo {title} {Influence of equilibrium shear flow along the magnetic field on the resistive tearing instability},}\ }\href {\doibase 10.1063/1.864061} {\bibfield  {journal} {\bibinfo  {journal} {Phys. Fluids}\ }\textbf {\bibinfo {volume} {26}},\ \bibinfo {pages} {2966--2975} (\bibinfo {year} {1983})}\BibitemShut {NoStop}%
\bibitem [{\citenamefont {Einaudi}\ and\ \citenamefont {Rubini}(1986)}]{Einaudi1986}%
  \BibitemOpen
  \bibfield  {author} {\bibinfo {author} {\bibfnamefont {G.}~\bibnamefont {Einaudi}}\ and\ \bibinfo {author} {\bibfnamefont {F.}~\bibnamefont {Rubini}},\ }\bibfield  {title} {\enquote {\bibinfo {title} {{Resistive instabilities in a flowing plasma: I. Inviscid case}},}\ }\href {\doibase 10.1063/1.865548} {\bibfield  {journal} {\bibinfo  {journal} {Phys. Fluids}\ }\textbf {\bibinfo {volume} {29}},\ \bibinfo {pages} {2563} (\bibinfo {year} {1986})}\BibitemShut {NoStop}%
\bibitem [{\citenamefont {Chen}\ and\ \citenamefont {Morrison}(1990)}]{Chen1990}%
  \BibitemOpen
  \bibfield  {author} {\bibinfo {author} {\bibfnamefont {X.~L.}\ \bibnamefont {Chen}}\ and\ \bibinfo {author} {\bibfnamefont {P.~J.}\ \bibnamefont {Morrison}},\ }\bibfield  {title} {\enquote {\bibinfo {title} {Resistive tearing instability with equilibrium shear flow},}\ }\href {\doibase 10.1063/1.859339} {\bibfield  {journal} {\bibinfo  {journal} {Phys. Fluids B: Plasma Phys.}\ }\textbf {\bibinfo {volume} {2}},\ \bibinfo {pages} {495--507} (\bibinfo {year} {1990})}\BibitemShut {NoStop}%
\bibitem [{\citenamefont {Zhang}\ \emph {et~al.}(2011)\citenamefont {Zhang}, \citenamefont {Li}, \citenamefont {Wang}, \citenamefont {Li},\ and\ \citenamefont {Ma}}]{Zhang2011}%
  \BibitemOpen
  \bibfield  {author} {\bibinfo {author} {\bibfnamefont {X.}~\bibnamefont {Zhang}}, \bibinfo {author} {\bibfnamefont {L.~J.}\ \bibnamefont {Li}}, \bibinfo {author} {\bibfnamefont {L.~C.}\ \bibnamefont {Wang}}, \bibinfo {author} {\bibfnamefont {J.~H.}\ \bibnamefont {Li}}, \ and\ \bibinfo {author} {\bibfnamefont {Z.~W.}\ \bibnamefont {Ma}},\ }\bibfield  {title} {\enquote {\bibinfo {title} {Influences of sub-{Alfv\'enic} shear flows on nonlinear evolution of magnetic reconnection in compressible plasmas},}\ }\href {\doibase 10.1063/1.3643792} {\bibfield  {journal} {\bibinfo  {journal} {Phys. Plasmas}\ }\textbf {\bibinfo {volume} {18}},\ \bibinfo {pages} {092112} (\bibinfo {year} {2011})}\BibitemShut {NoStop}%
\bibitem [{\citenamefont {Wu}\ and\ \citenamefont {Ma}(2014)}]{Wu2014}%
  \BibitemOpen
  \bibfield  {author} {\bibinfo {author} {\bibfnamefont {L.~N.}\ \bibnamefont {Wu}}\ and\ \bibinfo {author} {\bibfnamefont {Z.~W.}\ \bibnamefont {Ma}},\ }\bibfield  {title} {\enquote {\bibinfo {title} {Linear growth rates of resistive tearing modes with sub-{Alfv'enic} streaming flow},}\ }\href {\doibase 10.1063/1.4886360} {\bibfield  {journal} {\bibinfo  {journal} {Phys. Plasmas}\ }\textbf {\bibinfo {volume} {21}},\ \bibinfo {pages} {072105} (\bibinfo {year} {2014})}\BibitemShut {NoStop}%
\bibitem [{\citenamefont {{Shi}}(2022)}]{Shi2022}%
  \BibitemOpen
  \bibfield  {author} {\bibinfo {author} {\bibfnamefont {C.}~\bibnamefont {{Shi}}},\ }\bibfield  {title} {\enquote {\bibinfo {title} {{Instabilities in a current sheet with plasma jet}},}\ }\href {\doibase 10.1017/S0022377822000575} {\bibfield  {journal} {\bibinfo  {journal} {J. Plasma Phys.}\ }\textbf {\bibinfo {volume} {88}},\ \bibinfo {eid} {555880401} (\bibinfo {year} {2022})}\BibitemShut {NoStop}%
\bibitem [{\citenamefont {{Faganello}}\ \emph {et~al.}(2010)\citenamefont {{Faganello}}, \citenamefont {{Pegoraro}}, \citenamefont {{Califano}},\ and\ \citenamefont {{Marradi}}}]{Faganello2010}%
  \BibitemOpen
  \bibfield  {author} {\bibinfo {author} {\bibfnamefont {M.}~\bibnamefont {{Faganello}}}, \bibinfo {author} {\bibfnamefont {F.}~\bibnamefont {{Pegoraro}}}, \bibinfo {author} {\bibfnamefont {F.}~\bibnamefont {{Califano}}}, \ and\ \bibinfo {author} {\bibfnamefont {L.}~\bibnamefont {{Marradi}}},\ }\bibfield  {title} {\enquote {\bibinfo {title} {{Collisionless magnetic reconnection in the presence of a sheared velocity field}},}\ }\href {\doibase 10.1063/1.3430640} {\bibfield  {journal} {\bibinfo  {journal} {Phys. Plasmas}\ }\textbf {\bibinfo {volume} {17}},\ \bibinfo {eid} {062102} (\bibinfo {year} {2010})}\BibitemShut {NoStop}%
\bibitem [{\citenamefont {{Tassi}}, \citenamefont {{Grasso}},\ and\ \citenamefont {{Comisso}}(2014)}]{Tassi2014}%
  \BibitemOpen
  \bibfield  {author} {\bibinfo {author} {\bibfnamefont {E.}~\bibnamefont {{Tassi}}}, \bibinfo {author} {\bibfnamefont {D.}~\bibnamefont {{Grasso}}}, \ and\ \bibinfo {author} {\bibfnamefont {L.}~\bibnamefont {{Comisso}}},\ }\bibfield  {title} {\enquote {\bibinfo {title} {{Linear stability analysis of collisionless reconnection in the presence of an equilibrium flow aligned with the guide field}},}\ }\href {\doibase 10.1140/epjd/e2014-40730-6} {\bibfield  {journal} {\bibinfo  {journal} {Eur. Phys. J. D}\ }\textbf {\bibinfo {volume} {68}},\ \bibinfo {eid} {88} (\bibinfo {year} {2014})}\BibitemShut {NoStop}%
\bibitem [{\citenamefont {{Biskamp}}(2000)}]{Biskamp2000}%
  \BibitemOpen
  \bibfield  {author} {\bibinfo {author} {\bibfnamefont {D.}~\bibnamefont {{Biskamp}}},\ }\href@noop {} {\emph {\bibinfo {title} {{Magnetic Reconnection in Plasmas}}}},\ Vol.~\bibinfo {volume} {3}\ (\bibinfo {year} {2000})\BibitemShut {NoStop}%
\bibitem [{\citenamefont {{Loureiro}}, \citenamefont {{Schekochihin}},\ and\ \citenamefont {{Uzdensky}}(2013)}]{Loureiro2013}%
  \BibitemOpen
  \bibfield  {author} {\bibinfo {author} {\bibfnamefont {N.~F.}\ \bibnamefont {{Loureiro}}}, \bibinfo {author} {\bibfnamefont {A.~A.}\ \bibnamefont {{Schekochihin}}}, \ and\ \bibinfo {author} {\bibfnamefont {D.~A.}\ \bibnamefont {{Uzdensky}}},\ }\bibfield  {title} {\enquote {\bibinfo {title} {{Plasmoid and Kelvin-Helmholtz instabilities in Sweet-Parker current sheets}},}\ }\href {\doibase 10.1103/PhysRevE.87.013102} {\bibfield  {journal} {\bibinfo  {journal} {Phys. Rev. E}\ }\textbf {\bibinfo {volume} {87}},\ \bibinfo {eid} {013102} (\bibinfo {year} {2013})},\ \Eprint {http://arxiv.org/abs/1208.0966} {arXiv:1208.0966 [physics.plasm-ph]} \BibitemShut {NoStop}%
\bibitem [{\citenamefont {{Keppens}}\ \emph {et~al.}(1999)\citenamefont {{Keppens}}, \citenamefont {{T{\'o}th}}, \citenamefont {{Westermann}},\ and\ \citenamefont {{Goedbloed}}}]{Keppens1999}%
  \BibitemOpen
  \bibfield  {author} {\bibinfo {author} {\bibfnamefont {R.}~\bibnamefont {{Keppens}}}, \bibinfo {author} {\bibfnamefont {G.}~\bibnamefont {{T{\'o}th}}}, \bibinfo {author} {\bibfnamefont {R.~H.~J.}\ \bibnamefont {{Westermann}}}, \ and\ \bibinfo {author} {\bibfnamefont {J.~P.}\ \bibnamefont {{Goedbloed}}},\ }\bibfield  {title} {\enquote {\bibinfo {title} {{Growth and saturation of the Kelvin-Helmholtz instability with parallel and antiparallel magnetic fields}},}\ }\href {\doibase 10.1017/S0022377898007223} {\bibfield  {journal} {\bibinfo  {journal} {J. Plasma Phys.}\ }\textbf {\bibinfo {volume} {61}},\ \bibinfo {pages} {1--19} (\bibinfo {year} {1999})},\ \Eprint {http://arxiv.org/abs/astro-ph/9901166} {arXiv:astro-ph/9901166 [astro-ph]} \BibitemShut {NoStop}%
\bibitem [{\citenamefont {{Borgogno}}\ \emph {et~al.}(2022)\citenamefont {{Borgogno}}, \citenamefont {{Grasso}}, \citenamefont {{Achilli}}, \citenamefont {{Rom{\'e}}},\ and\ \citenamefont {{Comisso}}}]{Borgogno2022}%
  \BibitemOpen
  \bibfield  {author} {\bibinfo {author} {\bibfnamefont {D.}~\bibnamefont {{Borgogno}}}, \bibinfo {author} {\bibfnamefont {D.}~\bibnamefont {{Grasso}}}, \bibinfo {author} {\bibfnamefont {B.}~\bibnamefont {{Achilli}}}, \bibinfo {author} {\bibfnamefont {M.}~\bibnamefont {{Rom{\'e}}}}, \ and\ \bibinfo {author} {\bibfnamefont {L.}~\bibnamefont {{Comisso}}},\ }\bibfield  {title} {\enquote {\bibinfo {title} {{Coexistence of Plasmoid and Kelvin-Helmholtz Instabilities in Collisionless Plasma Turbulence}},}\ }\href {\doibase 10.3847/1538-4357/ac582f} {\bibfield  {journal} {\bibinfo  {journal} {Astrophys. J.}\ }\textbf {\bibinfo {volume} {929}},\ \bibinfo {eid} {62} (\bibinfo {year} {2022})}\BibitemShut {NoStop}%
\bibitem [{\citenamefont {Park}\ \emph {et~al.}(2013)\citenamefont {Park}, \citenamefont {Sabbagh}, \citenamefont {Bialek}, \citenamefont {Berkery}, \citenamefont {Lee}, \citenamefont {Ko}, \citenamefont {Bak}, \citenamefont {Jeon}, \citenamefont {Park}, \citenamefont {Kim}, \citenamefont {Hahn}, \citenamefont {Ahn}, \citenamefont {Yoon}, \citenamefont {Lee}, \citenamefont {Choi}, \citenamefont {Yun}, \citenamefont {Park}, \citenamefont {You}, \citenamefont {Bae}, \citenamefont {Oh}, \citenamefont {Kim},\ and\ \citenamefont {Kwak}}]{Park2013}%
  \BibitemOpen
  \bibfield  {author} {\bibinfo {author} {\bibfnamefont {Y.}~\bibnamefont {Park}}, \bibinfo {author} {\bibfnamefont {S.}~\bibnamefont {Sabbagh}}, \bibinfo {author} {\bibfnamefont {J.}~\bibnamefont {Bialek}}, \bibinfo {author} {\bibfnamefont {J.}~\bibnamefont {Berkery}}, \bibinfo {author} {\bibfnamefont {S.}~\bibnamefont {Lee}}, \bibinfo {author} {\bibfnamefont {W.}~\bibnamefont {Ko}}, \bibinfo {author} {\bibfnamefont {J.}~\bibnamefont {Bak}}, \bibinfo {author} {\bibfnamefont {Y.}~\bibnamefont {Jeon}}, \bibinfo {author} {\bibfnamefont {J.}~\bibnamefont {Park}}, \bibinfo {author} {\bibfnamefont {J.}~\bibnamefont {Kim}}, \bibinfo {author} {\bibfnamefont {S.}~\bibnamefont {Hahn}}, \bibinfo {author} {\bibfnamefont {J.-W.}\ \bibnamefont {Ahn}}, \bibinfo {author} {\bibfnamefont {S.}~\bibnamefont {Yoon}}, \bibinfo {author} {\bibfnamefont {K.}~\bibnamefont {Lee}}, \bibinfo {author} {\bibfnamefont {M.}~\bibnamefont {Choi}}, \bibinfo {author} {\bibfnamefont {G.}~\bibnamefont {Yun}}, \bibinfo {author} {\bibfnamefont {H.}~\bibnamefont {Park}}, \bibinfo {author} {\bibfnamefont {K.-I.}\ \bibnamefont {You}}, \bibinfo {author} {\bibfnamefont {Y.}~\bibnamefont {Bae}}, \bibinfo {author} {\bibfnamefont {Y.}~\bibnamefont {Oh}}, \bibinfo {author} {\bibfnamefont {W.-C.}\ \bibnamefont {Kim}}, \ and\ \bibinfo {author} {\bibfnamefont {J.}~\bibnamefont {Kwak}},\ }\bibfield  {title} {\enquote {\bibinfo {title} {Investigation of mhd instabilities and control in kstar preparing for high beta operation},}\ }\href {\doibase 10.1088/0029-5515/53/8/083029} {\bibfield  {journal} {\bibinfo  {journal} {Nucl. Fusion}\ }\textbf {\bibinfo {volume} {53}},\ \bibinfo {pages} {083029} (\bibinfo {year} {2013})}\BibitemShut {NoStop}%
\bibitem [{\citenamefont {Shao}\ \emph {et~al.}(2021)\citenamefont {Shao}, \citenamefont {Liu}, \citenamefont {Xu}, \citenamefont {Chen}, \citenamefont {Wang}, \citenamefont {Cheng}, \citenamefont {Wang}, \citenamefont {Huang}, \citenamefont {Liu}, \citenamefont {Zhang}, \citenamefont {Xu}, \citenamefont {Tang},\ and\ \citenamefont {Team}}]{Shao2021}%
  \BibitemOpen
  \bibfield  {author} {\bibinfo {author} {\bibfnamefont {J.}~\bibnamefont {Shao}}, \bibinfo {author} {\bibfnamefont {H.}~\bibnamefont {Liu}}, \bibinfo {author} {\bibfnamefont {Y.}~\bibnamefont {Xu}}, \bibinfo {author} {\bibfnamefont {Z.}~\bibnamefont {Chen}}, \bibinfo {author} {\bibfnamefont {T.}~\bibnamefont {Wang}}, \bibinfo {author} {\bibfnamefont {J.}~\bibnamefont {Cheng}}, \bibinfo {author} {\bibfnamefont {X.}~\bibnamefont {Wang}}, \bibinfo {author} {\bibfnamefont {J.}~\bibnamefont {Huang}}, \bibinfo {author} {\bibfnamefont {H.}~\bibnamefont {Liu}}, \bibinfo {author} {\bibfnamefont {X.}~\bibnamefont {Zhang}}, \bibinfo {author} {\bibfnamefont {K.}~\bibnamefont {Xu}}, \bibinfo {author} {\bibfnamefont {C.}~\bibnamefont {Tang}}, \ and\ \bibinfo {author} {\bibfnamefont {T.~J.-T.}\ \bibnamefont {Team}},\ }\bibfield  {title} {\enquote {\bibinfo {title} {Effect of the toroidal flow and flow shear on the m/n 2/1 tearing mode in j-text tokamak},}\ }\href {\doibase 10.1088/1361-6587/abf85e} {\bibfield  {journal} {\bibinfo  {journal} {Plasma Phys. and Control. Fusion}\ }\textbf {\bibinfo {volume} {63}},\ \bibinfo {pages} {065017} (\bibinfo {year} {2021})}\BibitemShut {NoStop}%
\bibitem [{\citenamefont {Chu}\ \emph {et~al.}(1995)\citenamefont {Chu}, \citenamefont {Greene}, \citenamefont {Jensen}, \citenamefont {Miller}, \citenamefont {Bondeson}, \citenamefont {Johnson},\ and\ \citenamefont {Mauel}}]{Chu1995}%
  \BibitemOpen
  \bibfield  {author} {\bibinfo {author} {\bibfnamefont {M.~S.}\ \bibnamefont {Chu}}, \bibinfo {author} {\bibfnamefont {J.~M.}\ \bibnamefont {Greene}}, \bibinfo {author} {\bibfnamefont {T.~H.}\ \bibnamefont {Jensen}}, \bibinfo {author} {\bibfnamefont {R.~L.}\ \bibnamefont {Miller}}, \bibinfo {author} {\bibfnamefont {A.}~\bibnamefont {Bondeson}}, \bibinfo {author} {\bibfnamefont {R.~W.}\ \bibnamefont {Johnson}}, \ and\ \bibinfo {author} {\bibfnamefont {M.~E.}\ \bibnamefont {Mauel}},\ }\bibfield  {title} {\enquote {\bibinfo {title} {{Effect of toroidal plasma flow and flow shear on global magnetohydrodynamic MHD modes}},}\ }\href {\doibase 10.1063/1.871247} {\bibfield  {journal} {\bibinfo  {journal} {Phys. Plasmas}\ }\textbf {\bibinfo {volume} {2}},\ \bibinfo {pages} {2236--2241} (\bibinfo {year} {1995})}\BibitemShut {NoStop}%
\bibitem [{\citenamefont {White}\ and\ \citenamefont {Fitzpatrick}(2015)}]{White2015}%
  \BibitemOpen
  \bibfield  {author} {\bibinfo {author} {\bibfnamefont {R.~L.}\ \bibnamefont {White}}\ and\ \bibinfo {author} {\bibfnamefont {R.}~\bibnamefont {Fitzpatrick}},\ }\bibfield  {title} {\enquote {\bibinfo {title} {{Effect of rotation and velocity shear on tearing layer stability in tokamak plasmas}},}\ }\href {\doibase 10.1063/1.4932994} {\bibfield  {journal} {\bibinfo  {journal} {Phys. Plasmas}\ }\textbf {\bibinfo {volume} {22}},\ \bibinfo {pages} {102507} (\bibinfo {year} {2015})}\BibitemShut {NoStop}%
\bibitem [{\citenamefont {Cai}\ and\ \citenamefont {Cao}(2018)}]{Cai2018}%
  \BibitemOpen
  \bibfield  {author} {\bibinfo {author} {\bibfnamefont {H.}~\bibnamefont {Cai}}\ and\ \bibinfo {author} {\bibfnamefont {J.}~\bibnamefont {Cao}},\ }\bibfield  {title} {\enquote {\bibinfo {title} {Influence of toroidal rotation on magnetic islands in tokamaks},}\ }\href {\doibase 10.1088/1741-4326/aaa55e} {\bibfield  {journal} {\bibinfo  {journal} {Nucl. Fusion}\ }\textbf {\bibinfo {volume} {58}},\ \bibinfo {pages} {036008} (\bibinfo {year} {2018})}\BibitemShut {NoStop}%
\bibitem [{\citenamefont {Chen}\ and\ \citenamefont {Morrison}(1992)}]{Chen1992}%
  \BibitemOpen
  \bibfield  {author} {\bibinfo {author} {\bibfnamefont {X.~L.}\ \bibnamefont {Chen}}\ and\ \bibinfo {author} {\bibfnamefont {P.~J.}\ \bibnamefont {Morrison}},\ }\bibfield  {title} {\enquote {\bibinfo {title} {{Nonlinear interactions of tearing modes in the presence of shear flow}},}\ }\href {\doibase 10.1063/1.860238} {\bibfield  {journal} {\bibinfo  {journal} {Phys. Fluids B: Plasma Phys.}\ }\textbf {\bibinfo {volume} {4}},\ \bibinfo {pages} {845--854} (\bibinfo {year} {1992})}\BibitemShut {NoStop}%
\bibitem [{\citenamefont {Smolyakov}\ \emph {et~al.}(2001)\citenamefont {Smolyakov}, \citenamefont {Lazzaro}, \citenamefont {Azumi},\ and\ \citenamefont {Kishimoto}}]{Smolyakov2001}%
  \BibitemOpen
  \bibfield  {author} {\bibinfo {author} {\bibfnamefont {A.~I.}\ \bibnamefont {Smolyakov}}, \bibinfo {author} {\bibfnamefont {E.}~\bibnamefont {Lazzaro}}, \bibinfo {author} {\bibfnamefont {M.}~\bibnamefont {Azumi}}, \ and\ \bibinfo {author} {\bibfnamefont {Y.}~\bibnamefont {Kishimoto}},\ }\bibfield  {title} {\enquote {\bibinfo {title} {Stabilization of magnetic islands due to the sheared plasma flow and viscosity},}\ }\href {\doibase 10.1088/0741-3335/43/12/303} {\bibfield  {journal} {\bibinfo  {journal} {Plasma Phys. Control. Fusion}\ }\textbf {\bibinfo {volume} {43}},\ \bibinfo {pages} {1661} (\bibinfo {year} {2001})}\BibitemShut {NoStop}%
\bibitem [{\citenamefont {Ren}\ \emph {et~al.}(2022)\citenamefont {Ren}, \citenamefont {Wang}, \citenamefont {Cai},\ and\ \citenamefont {Liu}}]{Ren2023}%
  \BibitemOpen
  \bibfield  {author} {\bibinfo {author} {\bibfnamefont {Z.}~\bibnamefont {Ren}}, \bibinfo {author} {\bibfnamefont {F.}~\bibnamefont {Wang}}, \bibinfo {author} {\bibfnamefont {H.}~\bibnamefont {Cai}}, \ and\ \bibinfo {author} {\bibfnamefont {J.}~\bibnamefont {Liu}},\ }\bibfield  {title} {\enquote {\bibinfo {title} {Influence of toroidal rotation on nonlinear evolution of tearing mode in tokamak plasmas},}\ }\href {\doibase 10.1088/1361-6587/aca4f4} {\bibfield  {journal} {\bibinfo  {journal} {Plasma Phys. Control. Fusion}\ }\textbf {\bibinfo {volume} {65}},\ \bibinfo {pages} {015007} (\bibinfo {year} {2022})}\BibitemShut {NoStop}%
\bibitem [{\citenamefont {Claes}, \citenamefont {{De Jonghe}},\ and\ \citenamefont {Keppens}(2020)}]{Claes2020}%
  \BibitemOpen
  \bibfield  {author} {\bibinfo {author} {\bibfnamefont {N.}~\bibnamefont {Claes}}, \bibinfo {author} {\bibfnamefont {J.}~\bibnamefont {{De Jonghe}}}, \ and\ \bibinfo {author} {\bibfnamefont {R.}~\bibnamefont {Keppens}},\ }\bibfield  {title} {\enquote {\bibinfo {title} {Legolas: A modern tool for magnetohydrodynamic spectroscopy},}\ }\href {\doibase 10.3847/1538-4365/abc5c4} {\bibfield  {journal} {\bibinfo  {journal} {Astrophys. J. Suppl. Ser.}\ }\textbf {\bibinfo {volume} {251}},\ \bibinfo {pages} {25} (\bibinfo {year} {2020})}\BibitemShut {NoStop}%
\bibitem [{\citenamefont {{Claes}}\ and\ \citenamefont {{Keppens}}(2023)}]{Claes2023}%
  \BibitemOpen
  \bibfield  {author} {\bibinfo {author} {\bibfnamefont {N.}~\bibnamefont {{Claes}}}\ and\ \bibinfo {author} {\bibfnamefont {R.}~\bibnamefont {{Keppens}}},\ }\bibfield  {title} {\enquote {\bibinfo {title} {{Legolas 2.0: Improvements and extensions to an MHD spectroscopic framework}},}\ }\href {\doibase 10.1016/j.cpc.2023.108856} {\bibfield  {journal} {\bibinfo  {journal} {Comput. Phys. Commun.}\ }\textbf {\bibinfo {volume} {291}},\ \bibinfo {eid} {108856} (\bibinfo {year} {2023})},\ \Eprint {http://arxiv.org/abs/2307.10145} {arXiv:2307.10145 [astro-ph.IM]} \BibitemShut {NoStop}%
\bibitem [{Note1()}]{Note1}%
  \BibitemOpen
  \bibinfo {note} {Sometimes, the temperature is chosen as the constant quantity and subsequently, the density profile is determined by Eq. (\ref {eq:equil-cond}) such as in e.g. Ref. \protect \rev@citealpnum {Goedbloed2019}, Sec. $14.4.1$.}\BibitemShut {Stop}%
\bibitem [{\citenamefont {Ofman}, \citenamefont {Morrison},\ and\ \citenamefont {Steinolfson}(1993)}]{Ofman1993}%
  \BibitemOpen
  \bibfield  {author} {\bibinfo {author} {\bibfnamefont {L.}~\bibnamefont {Ofman}}, \bibinfo {author} {\bibfnamefont {P.~J.}\ \bibnamefont {Morrison}}, \ and\ \bibinfo {author} {\bibfnamefont {R.~S.}\ \bibnamefont {Steinolfson}},\ }\bibfield  {title} {\enquote {\bibinfo {title} {{Nonlinear evolution of resistive tearing mode instability with shear flow and viscosity}},}\ }\href {\doibase 10.1063/1.860523} {\bibfield  {journal} {\bibinfo  {journal} {Phys. Fluids B}\ }\textbf {\bibinfo {volume} {5}},\ \bibinfo {pages} {376--387} (\bibinfo {year} {1993})}\BibitemShut {NoStop}%
\bibitem [{\citenamefont {{Cross}}\ and\ \citenamefont {{Van Hoven}}(1971)}]{Cross1971}%
  \BibitemOpen
  \bibfield  {author} {\bibinfo {author} {\bibfnamefont {M.~A.}\ \bibnamefont {{Cross}}}\ and\ \bibinfo {author} {\bibfnamefont {G.}~\bibnamefont {{Van Hoven}}},\ }\bibfield  {title} {\enquote {\bibinfo {title} {{Magnetic and Gravitational Energy Release by Resistive Instabilities}},}\ }\href {\doibase 10.1103/PhysRevA.4.2347} {\bibfield  {journal} {\bibinfo  {journal} {Phys. Rev. A}\ }\textbf {\bibinfo {volume} {4}},\ \bibinfo {pages} {2347--2353} (\bibinfo {year} {1971})}\BibitemShut {NoStop}%
\bibitem [{\citenamefont {{Van Hoven}}\ and\ \citenamefont {{Cross}}(1973)}]{VanHoven1973}%
  \BibitemOpen
  \bibfield  {author} {\bibinfo {author} {\bibfnamefont {G.}~\bibnamefont {{Van Hoven}}}\ and\ \bibinfo {author} {\bibfnamefont {M.~A.}\ \bibnamefont {{Cross}}},\ }\bibfield  {title} {\enquote {\bibinfo {title} {{Energy Release by Magnetic Tearing: The Nonlinear Limit}},}\ }\href {\doibase 10.1103/PhysRevA.7.1347} {\bibfield  {journal} {\bibinfo  {journal} {Phys. Rev. A}\ }\textbf {\bibinfo {volume} {7}},\ \bibinfo {pages} {1347--1352} (\bibinfo {year} {1973})}\BibitemShut {NoStop}%
\bibitem [{\citenamefont {Goedbloed}, \citenamefont {Keppens},\ and\ \citenamefont {Poedts}(2019)}]{Goedbloed2019}%
  \BibitemOpen
  \bibfield  {author} {\bibinfo {author} {\bibfnamefont {H.}~\bibnamefont {Goedbloed}}, \bibinfo {author} {\bibfnamefont {R.}~\bibnamefont {Keppens}}, \ and\ \bibinfo {author} {\bibfnamefont {S.}~\bibnamefont {Poedts}},\ }\href {\doibase 10.1017/9781316403679} {\emph {\bibinfo {title} {{Magnetohydrodynamics of Laboratory and Astrophysical Plasmas}}}}\ (\bibinfo  {publisher} {Cambridge University Press},\ \bibinfo {year} {2019})\BibitemShut {NoStop}%
\bibitem [{\citenamefont {{Betar}}\ \emph {et~al.}(2022)\citenamefont {{Betar}}, \citenamefont {{Del Sarto}}, \citenamefont {{Ottaviani}},\ and\ \citenamefont {{Ghizzo}}}]{Betar2022}%
  \BibitemOpen
  \bibfield  {author} {\bibinfo {author} {\bibfnamefont {H.}~\bibnamefont {{Betar}}}, \bibinfo {author} {\bibfnamefont {D.}~\bibnamefont {{Del Sarto}}}, \bibinfo {author} {\bibfnamefont {M.}~\bibnamefont {{Ottaviani}}}, \ and\ \bibinfo {author} {\bibfnamefont {A.}~\bibnamefont {{Ghizzo}}},\ }\bibfield  {title} {\enquote {\bibinfo {title} {{Microscopic scales of linear tearing modes: a tutorial on boundary layer theory for magnetic reconnection}},}\ }\href {\doibase 10.1017/S0022377822001088} {\bibfield  {journal} {\bibinfo  {journal} {J. Plasma Phys.}\ }\textbf {\bibinfo {volume} {88}},\ \bibinfo {eid} {925880601} (\bibinfo {year} {2022})}\BibitemShut {NoStop}%
\bibitem [{\citenamefont {{Furth}}, \citenamefont {{Rutherford}},\ and\ \citenamefont {{Selberg}}(1973)}]{Furth1973}%
  \BibitemOpen
  \bibfield  {author} {\bibinfo {author} {\bibfnamefont {H.~P.}\ \bibnamefont {{Furth}}}, \bibinfo {author} {\bibfnamefont {P.~H.}\ \bibnamefont {{Rutherford}}}, \ and\ \bibinfo {author} {\bibfnamefont {H.}~\bibnamefont {{Selberg}}},\ }\bibfield  {title} {\enquote {\bibinfo {title} {{Tearing mode in the cylindrical tokamak}},}\ }\href {\doibase 10.1063/1.1694467} {\bibfield  {journal} {\bibinfo  {journal} {Phys. Fluids}\ }\textbf {\bibinfo {volume} {16}},\ \bibinfo {pages} {1054--1063} (\bibinfo {year} {1973})}\BibitemShut {NoStop}%
\bibitem [{\citenamefont {{Pritchett}}, \citenamefont {{Lee}},\ and\ \citenamefont {{Drake}}(1980)}]{Pritchett1980}%
  \BibitemOpen
  \bibfield  {author} {\bibinfo {author} {\bibfnamefont {P.~L.}\ \bibnamefont {{Pritchett}}}, \bibinfo {author} {\bibfnamefont {Y.~C.}\ \bibnamefont {{Lee}}}, \ and\ \bibinfo {author} {\bibfnamefont {J.~F.}\ \bibnamefont {{Drake}}},\ }\bibfield  {title} {\enquote {\bibinfo {title} {{Linear analysis of the double-tearing mode}},}\ }\href {\doibase 10.1063/1.863151} {\bibfield  {journal} {\bibinfo  {journal} {Phys. Fluids}\ }\textbf {\bibinfo {volume} {23}},\ \bibinfo {pages} {1368--1374} (\bibinfo {year} {1980})}\BibitemShut {NoStop}%
\bibitem [{\citenamefont {{Betar}}\ \emph {et~al.}(2020)\citenamefont {{Betar}}, \citenamefont {{Del Sarto}}, \citenamefont {{Ottaviani}},\ and\ \citenamefont {{Ghizzo}}}]{Betar2020}%
  \BibitemOpen
  \bibfield  {author} {\bibinfo {author} {\bibfnamefont {H.}~\bibnamefont {{Betar}}}, \bibinfo {author} {\bibfnamefont {D.}~\bibnamefont {{Del Sarto}}}, \bibinfo {author} {\bibfnamefont {M.}~\bibnamefont {{Ottaviani}}}, \ and\ \bibinfo {author} {\bibfnamefont {A.}~\bibnamefont {{Ghizzo}}},\ }\bibfield  {title} {\enquote {\bibinfo {title} {{Multiparametric study of tearing modes in thin current sheets}},}\ }\href {\doibase 10.1063/5.0022133} {\bibfield  {journal} {\bibinfo  {journal} {Phys. Plasmas}\ }\textbf {\bibinfo {volume} {27}},\ \bibinfo {eid} {102106} (\bibinfo {year} {2020})}\BibitemShut {NoStop}%
\bibitem [{\citenamefont {Coppi}, \citenamefont {Greene},\ and\ \citenamefont {Johnson}(1966)}]{Coppi1966}%
  \BibitemOpen
  \bibfield  {author} {\bibinfo {author} {\bibfnamefont {B.}~\bibnamefont {Coppi}}, \bibinfo {author} {\bibfnamefont {J.~M.}\ \bibnamefont {Greene}}, \ and\ \bibinfo {author} {\bibfnamefont {J.~L.}\ \bibnamefont {Johnson}},\ }\bibfield  {title} {\enquote {\bibinfo {title} {Resistive instabilities in a diffuse linear pinch},}\ }\href {\doibase 10.1088/0029-5515/6/2/003} {\bibfield  {journal} {\bibinfo  {journal} {Nucl. Fusion}\ }\textbf {\bibinfo {volume} {6}},\ \bibinfo {pages} {101--117} (\bibinfo {year} {1966})}\BibitemShut {NoStop}%
\bibitem [{\citenamefont {Quarteroni}(2009)}]{Quarteroni2009}%
  \BibitemOpen
  \bibfield  {author} {\bibinfo {author} {\bibfnamefont {A.}~\bibnamefont {Quarteroni}},\ }\href@noop {} {\emph {\bibinfo {title} {Numerical Models for Differential Problems}}},\ Vol.~\bibinfo {volume} {2}\ (\bibinfo  {publisher} {Springer-Verlag},\ \bibinfo {year} {2009})\BibitemShut {NoStop}%
\end{thebibliography}%

\end{document}